\documentclass[aps,prl,reprint,floatfix,superscriptaddress]{revtex4-2}
\usepackage{xcolor}
\usepackage[utf8]{inputenc}
\usepackage{amsmath,amssymb,bm,graphicx,xcolor}
\usepackage[colorlinks=true,linkcolor=blue,citecolor=blue,urlcolor=blue]{hyperref}
\graphicspath{{figures/}}
\begin{document}
% V4 title: Discrete Time Crystals Stabilized by Dynamical Freezing in the Thermodynamic Limit
% Alternative: Anti-Unitary Floquet Driving Stabilizes Discrete Time Crystals in the Thermodynamic Limit
% (PRL asks authors not to introduce new terminology in titles, so "anti-unitary time crystal" is avoided.)
\title{Discrete Time Crystals from Anti-Unitary Floquet Driving and Dynamical Freezing}
\author{Anirban Das}
\affiliation{Indian Association for the Cultivation of Science (School of Physical Sciences), P.O. Jadavpur University, 2A \& 2B Raja S. C. Mullick Road, Kolkata 700032, West Bengal, India}
\author{Asmi Haldar}
\affiliation{Max Planck Institute for the Physics of Complex Systems, N\"othnitzer Stra\ss e 38, Dresden 01187, Germany}
\affiliation{Laboratoire de Physique Th\'eorique--IRSAMC, Paul Sabatier University, 118 route de Narbonne, building 3R1B4, 31400 Toulouse, France}
\author{Arnab Das}
\affiliation{Indian Association for the Cultivation of Science (School of Physical Sciences), P.O. Jadavpur University, 2A \& 2B Raja S. C. Mullick Road, Kolkata 700032, West Bengal, India}
\date{October 12, 2026}

\begin{abstract}
% 563 characters including spaces (PRL limit: 600).
We show anti-unitary periodic drive paves a direct path for stabilizing a Discrete Time-Crystal by Dynamical Freezing (DF). The magnetization appears as an emergent conserved operator, and an exact pairing of two-cycle quasi-energies pins the subharmonic response. DF keeps the effective Hamiltonian \textit{quasi-local}; infinite-chain simulations show no detectable decay for a generic state with any finite magnetization. We outline an experimental realization of the anti-unitary drive.
\end{abstract}
\maketitle

\noindent\textit{\textbf{Introduction:}}
Time crystals were proposed as phases of matter that spontaneously break continuous time-translation
symmetry~\cite{Wilczek2012}, but no-go theorems exclude this in ground states and thermal-equilibrium states of time-independent many-body Hamiltonians~\cite{Bruno2013,WatanabeOshikawa2015}. Periodically driven systems can instead break discrete time-translation symmetry and show a robust subharmonic response, the defining property of discrete time crystals (DTCs)~\cite{Sacha1,Khemani2016,Else2016,Zhang2017,Yao2017}. Stable DTCs require suppressing Floquet thermalization, by many-body localization in disordered systems~\cite{Khemani2016,Else2016,Yao2017,Zhang2017} or by prethermalization in clean ones~\cite{Else2017,Luitz2020}; see Refs.~\cite{SachaReview2018,KhemaniHistory2019,ZaletelReview2023} for reviews.

Clean, short-range one-dimensional (1D) chains are the most demanding setting, and the one in which infinite-system tensor networks access the thermodynamic limit directly. Prethermal DTCs based on spontaneously breaking an emergent $\mathbb{Z}_2$ symmetry cannot exist there at finite energy density~\cite{Else2017,Machado2020}. Period doubling can instead be carried by an approximately conserved magnetization~\cite{Luitz2020}, but that protection requires drive frequencies large compared with the local energy scales. Dynamical freezing (DF) works in a strong-drive regime instead~\cite{AD-DMF,AD-SDG,Mahesh_Freezing,Kris-Periodic,Russomanno_Dynamical_Freezing,Onset,Asmi_DF_PRX_2021,Adhip_Diptiman,Bhaskar_Scar,Bhaskar_DF,Roopayan_DF,Naveen_Dynamical_Freezing,Diptiman_DF,Somsubhra_Fragmentation_DF,Tista_Non-Hermitian_DF,Debanjan_DF_QDot,Debanjan_DF_Transmon,DF_TDL_iTEBD,Debanjan_FlowEq_DF,Debanjan_Frozonium_DF,DF_Lei,DF_Krish_Tapan,Lu2026_DF_Magnetometry}: strong periodic driving generates emergent conserved operators (ECOs) that retain memory of their initial expectation values although they do not commute with the instantaneous Hamiltonian~\cite{Onset,Asmi_DF_PRX_2021,DF_engineering_2026}. DF can be probed directly in the thermodynamic limit~\cite{DF_TDL_iTEBD}, and freezing points have been used to slow the decay of a kicked-Ising DTC~\cite{Collura2022}.

Here we show that inserting time reversal into the drive cycle turns DF into a route to robust time-crystalline order in a clean, non-integrable 1D chain. Complementing a global spin flip with complex conjugation makes the single-cycle Floquet map anti-unitary, with two consequences. First, the flip then echoes away the interactions rather than the strong field, so the magnetization, the very quantity the flip reverses, becomes the ECO. A conventional unitary flip protects only flip-even bond operators, and its period doubling is limited to special initial states and observables. Second, anti-unitarity enforces an exact pairing of two-cycle quasi-energies, $\varepsilon\leftrightarrow-\varepsilon$, which pins the subharmonic frequency independently of microscopic details and of errors in the flip angle. DF supplies the remaining ingredient, a local \cite{DF_engineering_2026}, non-heating effective Hamiltonian, and Floquet--Dyson perturbation theory (FDPT) at leading order shows that it does so through multiphoton interference rather than a static gap, failing only at isolated resonances. Infinite-chain iTEBD simulations show no detectable decay of the period-doubled response over accessible times and demonstrate its rigidity under variation of the longitudinal field. A single-ancilla embedding realizes the anti-unitary drive as a unitary circuit with the same period.\\

\begin{figure*}[tp]
\centering
\includegraphics[width=0.43\textwidth]{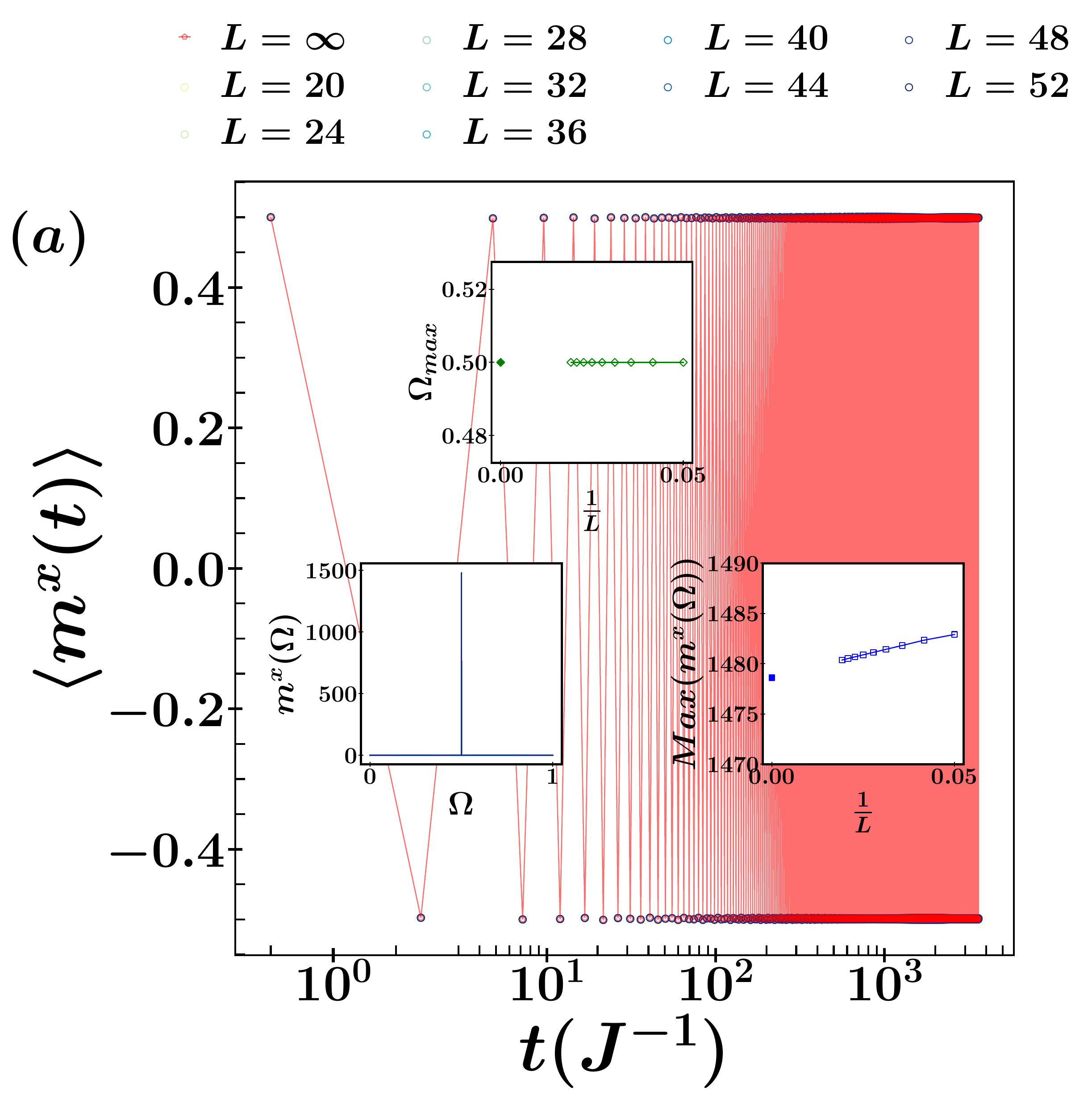}\hfill
\includegraphics[width=0.43\textwidth]{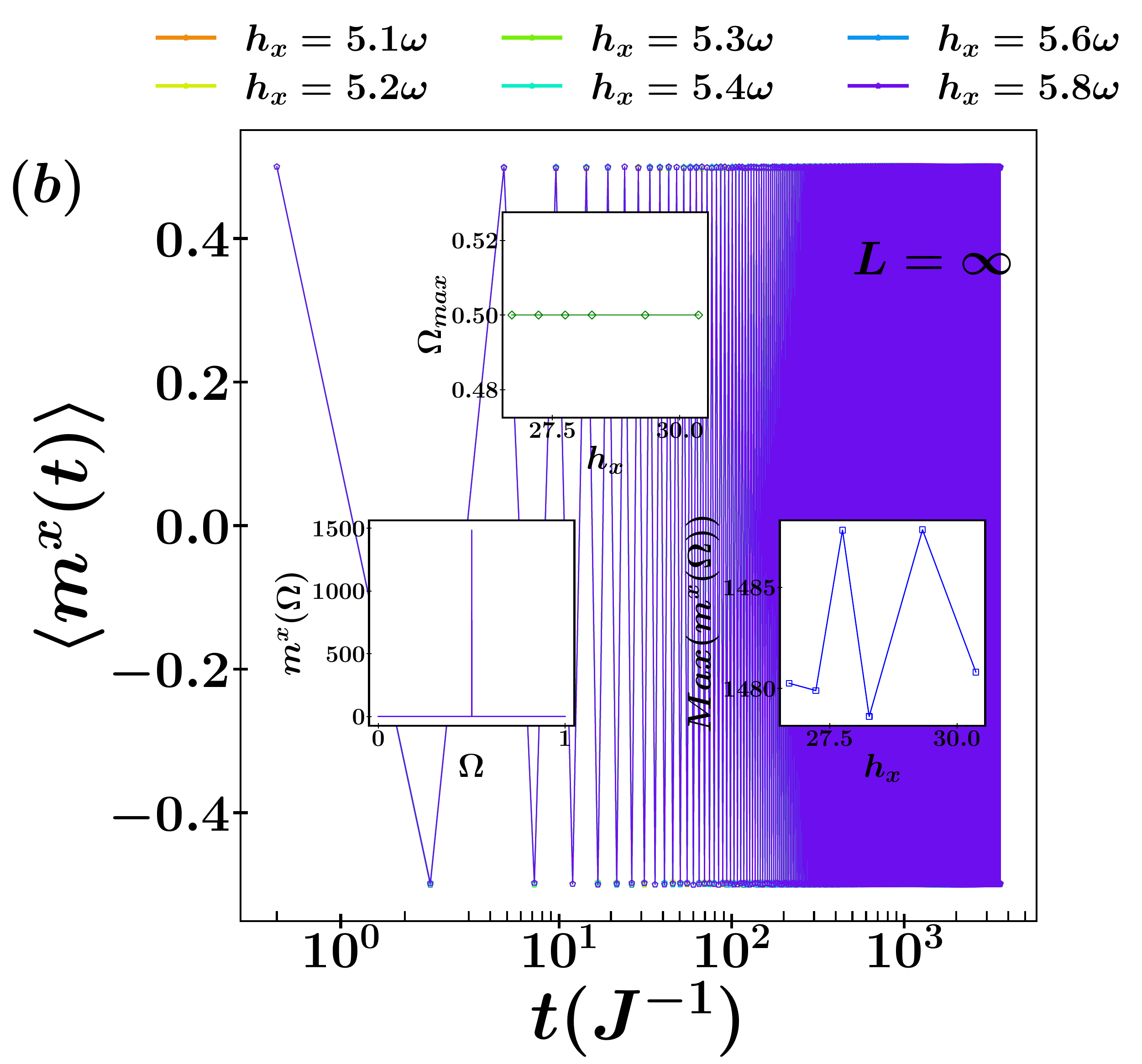}
\caption{{\bf Anti-unitary period doubling in the infinite chain.} {\bf (a)} Stroboscopic $m^x$ under $A_F$ from $|\uparrow\uparrow\downarrow\uparrow\cdots\rangle_x$ at $h_x=5.4\omega$: TEBD for $L=20$ to $52$ and iTEBD for $L=\infty$. Insets: Fourier spectrum (bottom left), dominant peak position $\Omega_{\max}$ versus $1/L$ (top), and peak strength at $\Omega=0.5$ versus $L$ (bottom right). {\bf (b)} Rigidity test: the same at $L=\infty$ for several $h_x$; the insets show the Fourier spectrum, and $\Omega_{\max}$ and the peak strength versus $h_x$. Throughout, $J=2$, $\kappa=1.7$, $\epsilon=\sqrt{3}/1.5$, and $\omega=5\pi/3$ were considered.}
\label{Fig:AU_dynamics}
\end{figure*}

% \begin{figure*}[tp]
% \centering
% \includegraphics[width=0.43\textwidth]{Fig1a.pdf}\hfill
% \includegraphics[width=0.43\textwidth]{Fig1b.pdf}
% \caption{{\bf Unitary period doubling in the infinite chain.} {\bf (a)} Stroboscopic $m^x$ under $U_F$ from $|\uparrow\uparrow\uparrow\cdots\rangle_x$ at $h_x=6\omega$ for several $L$ and $L=\infty$. Insets: Fourier spectrum (bottom left), $\Omega_{\max}$ versus $1/L$ with $L=\infty$ as a solid marker (top), and peak strength at $\Omega=0.5$ versus $L$ (bottom right). {\bf (b)} Rigidity test: the same at $L=\infty$ for several $h_x$, with insets and parameters as in Fig.~\ref{Fig:AU_dynamics}(b).}
% \label{Fig:U_dynamics}
% \end{figure*}

%\begin{center}

%\end{center}
\noindent
\textit{\textbf{Model and Protocols:}}
We study $H=H_{\rm int}+H_h+V$, where
\begin{equation}
H_{\rm int}=-J\sum_i\sigma^x_i\sigma^x_{i+1}+\kappa\sum_i\sigma^x_i\sigma^x_{i+2},
\label{Eq:Hamiltonian_1_V4}
\end{equation}
$H_h=-h_x\sum_i\sigma^x_i$, and $V=-\epsilon\sum_i\sigma^z_i$. $H_{\rm int}$ and $H_h$ commute, $V$ flips spins, and $H$ is non-integrable for generic couplings. Each cycle evolves the chain under $H$ for $T_1=2\pi/\omega$ and then applies the global flip $\mathcal{R}=\exp(-i\pi\sum_jS^y_j)$, with $S^y_j=\sigma^y_j/2$. We compare the unitary and anti-unitary single-cycle maps
\begin{equation}
U_F=\mathcal{R}\,e^{-iHT_1},\qquad A_F=\mathcal{R}\,\hat K e^{-iHT_1},
\label{Eq:protocols}
\end{equation}
where $\hat K$ complex conjugates wave-function coefficients in the $\sigma^z$ basis. $A_F$ is defined at the wave-function level but is realized exactly by a unitary, $T_1$-periodic drive on the chain plus one ancilla qubit~\cite{Casanova2011,Zhang2015_Embedding}. We track $m^x=L^{-1}\sum_i\sigma^x_i$; period doubling appears as a Fourier peak at $\Omega=0.5$ (in cycles$^{-1}$).\\
% We track $m^x=L^{-1}\sum_i\sigma^x_i$ and $C^x_r=L^{-1}\sum_i\sigma^x_i\sigma^x_{i+r}$ ($r=1,2$); period doubling appears as a Fourier peak at $\Omega=0.5$ (in cycles$^{-1}$).\\

%\begin{center}

%\end{center}

\noindent
\textit{\textbf{Role of Anti-Unitarity:}}
\textit{Echo:} $\mathcal{R}$ leaves $H_{\rm int}$ invariant and reverses $H_h+V$. Since $H$ and $\mathcal{R}$ are real in the $\sigma^z$ basis, $\hat Ke^{-iHT_1}\hat K=e^{+iHT_1}$: the conjugation reverses the direction of time in the following interval. Up to a global phase,
\begin{align}
U_F^2&=e^{-iT_1(H_{\rm int}-H_h-V)}\,e^{-iT_1(H_{\rm int}+H_h+V)},\nonumber\\
A_F^2&=e^{-iT_1(-H_{\rm int}+H_h+V)}\,e^{-iT_1(H_{\rm int}+H_h+V)}.
\label{eq:twocycle}
\end{align}
The flip thus acts as an echo, and $\hat K$ selects what it cancels. At $\epsilon=0$, where all terms commute, Eq.~\eqref{eq:twocycle} reduces exactly to
\begin{equation}
U_F^2\big|_{\epsilon=0}=e^{-2iT_1H_{\rm int}},\qquad A_F^2\big|_{\epsilon=0}=e^{-2iT_1H_h}.
\label{eq:echo}
\end{equation}
Under $U_F$, the field reverses between intervals, as in a standard DF drive of period $2T_1$. It is echoed away, the interactions survive, and the natural ECOs are the flip-even $C^x_{1,2}$. Under $A_F$, the interactions are echoed away, and the strong field survives, so the ECO is the flip-odd $m^x$ itself, precisely the quantity the flip reverses. A conserved flip-odd ECO toggles every cycle for any initial state with $m^x\neq0$. Flip-even ECOs constrain $m^x$ only within ECO-resolved sectors. The sector of the polarized state contains only that state and its flipped partner, but a generic sector contains states with different $m^x$, whose mixing adds frequencies to the response.

\par\smallskip
\noindent\textit{Antiunitary Pairing:} Anti-unitarity also constrains the response independently of the details of $H$. Let $A_F^2=e^{-2iT_1H_F}$. Since $A_F$ commutes with $A_F^2$ and is anti-unitary, $A_Fe^{-2iT_1H_F}A_F^{-1}=e^{2iT_1A_FH_FA_F^{-1}}=e^{-2iT_1H_F}$. The two-cycle quasi-energies therefore pair exactly, $\{\varepsilon_n\}=\{-\varepsilon_n\}$, and $A_F$ maps the Floquet state at $\varepsilon_n$ to the one at $-\varepsilon_n$. For the local generator connected continuously to $H_F|_{\epsilon=0}=H_h$,
\begin{equation}
A_F\,H_F\,A_F^{-1}=-H_F .
\label{eq:pairing}
\end{equation}
% Hence $\langle H_F\rangle$, approximately $-h_xL\langle m^x\rangle$, changes sign exactly every cycle, for any initial state, and the subharmonic frequency is fixed at $\Omega=1/2$ by symmetry. This holds for any perturbation that preserves the anti-unitarity of the cycle, including changes of couplings and fields and errors in the flip angle, as long as no resonance is crossed, so that $H_F$ stays local.

% An imperfect rotation $\mathcal{R}_\delta=\exp[-i(\pi-\delta)\sum_jS^y_j]$ is still real in the $\sigma^z$ basis and keeps the cycle anti-unitary, so the pairing survives it. 

Hence $\langle H_F\rangle$, approximately $-h_xL\langle m^x\rangle$, changes sign exactly every cycle, for any initial state, and the subharmonic frequency is fixed at $\Omega=1/2$ by symmetry. This holds for any perturbation that preserves the anti-unitarity of the cycle, including changes of couplings and fields, as long as no resonance is crossed, so that $H_F$ stays local. Equation~\eqref{eq:pairing} is the anti-unitary counterpart of the $\pi$-pairing of unitary DTCs~\cite{Khemani2016,Else2016,Yao2017}. There the pairing emerges dynamically; here it is enforced, and DF is needed only to keep $H_F$ local. Unlike the quasi-energy pairing produced by a unitary time reflection in Floquet supersymmetry~\cite{Iadecola2018}, it follows from the time step itself. No such constraint exists for $U_F$, which commutes with the generator of $U_F^2$.\\
%% [ED-CHECK] Two-cycle quasi-energies, L = 8, h_x = 5.4 omega:
%%   max |sort(eps) + sort(-eps)| = 6e-15 (anti-unitary) vs 0.64 (unitary).
%% [ED-CHECK, flip-angle errors; re-run for V6] Same system, R_delta = exp[-i(pi - delta) sum S^y] (real).
%%   Hausdorff distance between the spectrum of the two-cycle map and its complex conjugate (0 = exact pairing):
%%   delta = 0, 0.3, 0.5: <= 2e-14 (anti-unitary, A^2 = R_delta e^{+iHT} R_delta e^{-iHT}) vs 0.62, 0.26, 0.21 (unitary).
%% [ED-CHECK] 2048 cycles: anti-unitary (uudu, 5.4 omega, L = 12) peak at 0.5000 for delta = 0.1, 0.3, 0.5
%%   (weight near 1/2: 0.96, 0.84, 0.63); unitary (polarized, 6 omega, L = 8 and 12) peak 0.4966 at delta = 0.3.
%% The flip-angle statement in the pairing paragraph is theory only; none of these checks is cited in the text.
%%%%%%%%%%%%%%%%%%%%%%%%%%%%%%%%%%%%%%%%%%%%%%%%%
\begin{figure}[tp]
\centering
\includegraphics[width=0.75\linewidth]{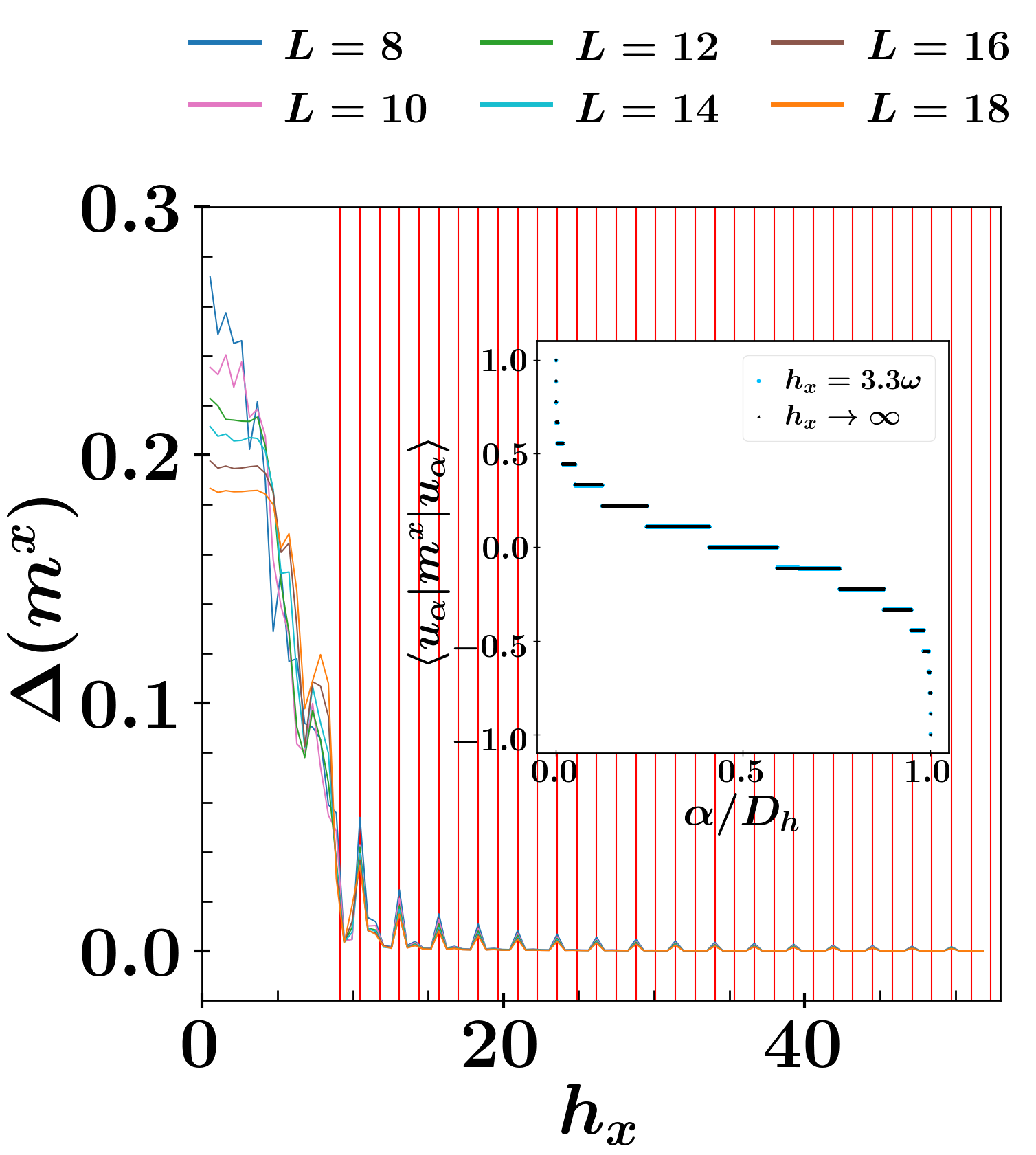}
% \caption{{\bf Emergent conservation of $m^x$ and its resonances.} $\Delta(m^x)$ under $A_F^2$ versus $h_x$ for $L=8$ to $18$. Red lines mark the first-order resonances of Eq.~\eqref{Resonance_Anti-Unitary}. Inset: ordered Floquet expectation values of $m^x$ and its eigenvalues at $h_x=3.3\omega$.}
\caption{\textbf{$m^x$ as an ECO in Anti-Unitary protocol.}
The spectral deviation $\Delta(m^x)$ is plotted against the
longitudinal field $h_x$ for $L=8$--$18$. It is computed from the
expectation values of $m^x$ in the Floquet eigenstates of the
two-cycle unitary operator $A_F^2$. At sufficiently large $h_x$,
$\Delta(m^x)$ remains small away from isolated resonances, indicating
approximate conservation of $m^x$ over two cycles. The red lines mark
the first-order resonance locations predicted by
Eq.~\eqref{Resonance_Anti-Unitary}. \textbf{Inset:} Ordered Floquet eigen expectation values of
$m^x$ compared with its ordered eigenvalues at $h_x=3.3\omega$.}
\label{Fig:1:ECO_Anti-Unitary}
\end{figure}
%%%%%%%%%%%%%%%%%%%%%%%%%%%%%%%%%%%%%%%%%%%%%%%%%
%
%%%%%%%%%%%%%%%%%%%%%%%%%%%%%%%%%%%%%%%%%%%%%%%%%
\begin{figure}[tp]
\centering
\includegraphics[width=0.42\textwidth]{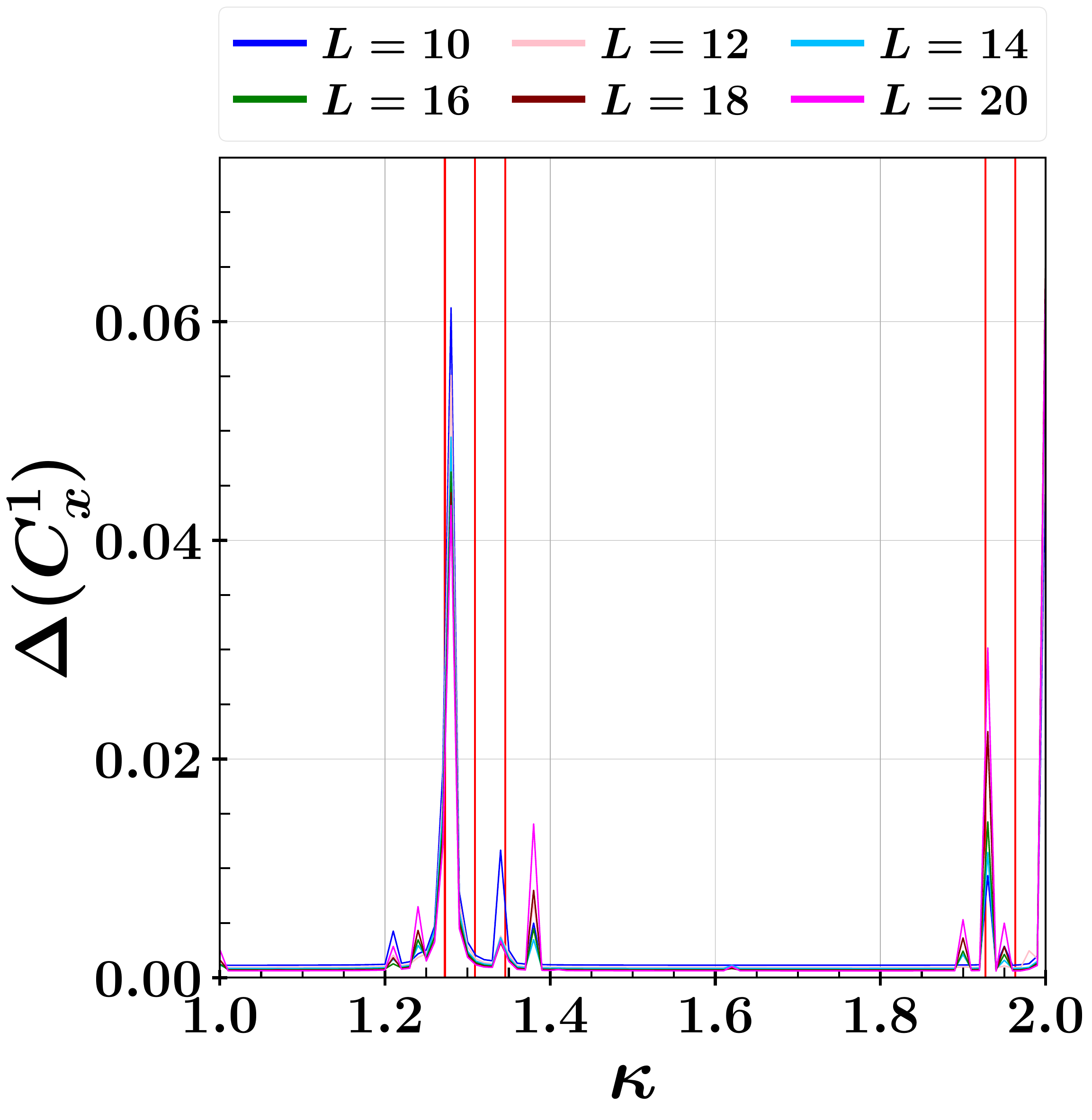}
\caption{{\bf ECO in Unitary setup and its resonances.} $\Delta(C^x_1)$ under $U_F$ versus $\kappa$ at $h_x=6\omega$ for several $L$. Red lines mark the first-order resonances of Eq.~\eqref{Resonances_Unitary}.}
\label{Fig:Resonances_Unitary_ECO}
\end{figure}
%%%%%%%%%%%%%%%%%%%%%%%%%%%%%%%%%%%
%
%%%%%%%%%%%%%%%%%%%%%%%%%%%%%%%%%%%
%\begin{center}

\noindent
\textit{\textbf{Thermodynamic Limit:}}
%\end{center}
Figure~\ref{Fig:AU_dynamics}(a) shows anti-unitary dynamics from the product state built from the four-site cell $|\uparrow\uparrow\downarrow\uparrow\rangle_x$, with $m^x=1/2$, at $h_x=5.4\omega$, away from the resonances discussed below. Finite-chain TEBD and infinite-chain iTEBD agree. The response shows no detectable decay over the accessible window; the dominant Fourier peak stays locked at $\Omega=0.5$ for all $L$ including $L=\infty$, and the peak strength is nonzero at $L=\infty$ and close to its finite-$L$ values. Figure~\ref{Fig:AU_dynamics}(b) tests rigidity by varying $h_x$, and hence $\epsilon/h_x$, directly at $L=\infty$: the peak stays locked at $\Omega=0.5$ and its strength shows no systematic drift, so the conservation protecting the response is not fine-tuned to a single value of $h_x$. The same holds for other $x$-basis product states with nonzero magnetization, including the polarized state at $L=\infty$, and for superposition states in finite chains at large $h_x$, as shown in the Supplemental Material (SM).

For comparison, the unitary protocol yields period doubling from the fully polarized state, with the dominant magnetization peak locked at $\Omega=0.5$ even at $L=\infty$. The response depends on the initial state and observable: period doubling appears in the staggered and four-site magnetizations for the N\'eel and $|\uparrow\uparrow\downarrow\downarrow\rangle_x$ states, respectively, while $m^x$ can show additional frequencies for other $x$-basis initial states (see SM).\\

\noindent
\textbf{\textit{Freezing Mechanism and Resonances}:} We identify ECOs spectrally. For an operator $\mathcal{O}$, we order its expectation values $o_n$ in the Floquet eigenstates of $U_F$, or of the unitary $A_F^2$, and its eigenvalues $\lambda_n(\mathcal{O})$, and define
\begin{equation}
\Delta(\mathcal{O})=\frac{1}{D_H}\sum_{n=1}^{D_H}\left|o_n-\lambda_n(\mathcal{O})\right|,
\label{spectral_deviation}
\end{equation}
with $D_H$ the Hilbert-space dimension. A small $\Delta$ that decreases with $L$ signals approximate conservation. Above a threshold field, this holds for $C^x_{1,2}$ under $U_F$ (Fig.~\ref{Fig:Resonances_Unitary_ECO} and SM) and for $m^x$ under $A_F^2$ (Fig.~\ref{Fig:1:ECO_Anti-Unitary}), except near isolated parameter values. As in other dynamically frozen systems, the ECOs deteriorate at isolated resonances~\cite{Asmi_DF_PRX_2021}.

\par\smallskip
\noindent\textit{Freezing, not a static gap:} By Eq.~\eqref{eq:twocycle}, $A_F^2$ contains a static strong field, so one might attribute the conservation of $m^x$ to the Zeeman gap $2h_x$. Although the bare Zeeman cost of a single spin flip, $2h_x$, is large, the two-cycle quasienergies are defined modulo $\omega/2$. The folded separation is therefore at most $\omega/4\approx1.3$, below $J=2$ and comparable to $\epsilon = \sqrt{3}/1.5$. 
%That gap is not available. Over two cycles, energy is conserved only modulo $\omega/2$, so a single-flip cost $2h_x$ folds to a quasi-energy detuning of at most $\omega/4\approx1.3$, below $J$ and comparable to $\epsilon\approx1.15$. 
A gap-based estimate would predict an $O(1)$ admixture, $\epsilon/(\omega/4)$, whereas the actual admixture is $\epsilon/(2h_x)\approx0.02$. To first order in FDPT (Supplemental Matter), flipping spin $\sigma_0$, with Zeeman cost $Z=2h_x\sigma_0$ and interaction cost $D=2[J\sigma_0(\sigma_1+\sigma_{-1})-\kappa\sigma_0(\sigma_2+\sigma_{-2})]$, gives the admixture
\begin{equation}
c=\frac{\epsilon}{Z-D}-\frac{2\epsilon D}{Z^2-D^2}\,\frac{e^{i(Z+D)T_1}-1}{e^{i\theta}-1},
\label{eq:admix}
\end{equation}
with $\theta=2ZT_1$.The leading contribution to the amplitude accumulated over one
two-step Floquet period is proportional to $e^{i\theta}-1$,
which cancels the corresponding factor in the FDPT denominator. This is the repeated-interference mechanism underlying DF~\cite{AD-DMF,AD-SDG,Mahesh_Freezing,Asmi_DF_PRX_2021}. Hence (i) off resonance, $|c|\simeq\epsilon/(2h_x)$, independent of $\omega$ and of the folded detuning, so conservation holds over frequency intervals rather than only at freezing peaks; (ii) resonances require $\theta\in2\pi\mathbb{Z}$, i.e.,
\begin{equation}
4h_x\sigma_0=q\omega,\qquad q\in\mathbb{Z},
\label{Resonance_Anti-Unitary}
\end{equation}
with widths in $h_x/\omega$ of order $|D|/(4\pi h_x)$, so the resonant fraction of parameter space vanishes as $h_x\to\infty$; and (iii) the same expression with $\theta=2DT_1$ governs $U_F$, giving resonances at
\begin{equation}
J\sigma_0(\sigma_1+\sigma_{-1})-\kappa\sigma_0(\sigma_2+\sigma_{-2})=\frac{p\omega}{4},\qquad p\in\mathbb{Z}.
\label{Resonances_Unitary}
\end{equation}
% V3 author check (carried over): For the stated y-axis pulse, V changes sign between the two intervals. At h_x=6 omega, the first-order single-flip numerator vanishes at even-p degeneracies. Check the SM numerator and the red-line interpretation.
% V5 note: Eq. (admix) confirms this. At h_x = 6 omega, Z T_1 is a multiple of 2 pi, so at theta = 2 pi p the
% resonant numerator is (-1)^p - 1; only odd-p lines of Eq. (Resonances_Unitary) are first-order resonances there.
Both protocols thus share one freezing mechanism. Anti-unitarity changes only which energy sets $\theta$, from the configuration-dependent $D$ to the uniform Zeeman cost $Z$, so that $h_x/\omega$ alone fixes the anti-unitary resonances. The predicted resonances in both the cases coincide with the peaks of $\Delta(m^x)$~(Anti-Unitary) in Fig.~\ref{Fig:1:ECO_Anti-Unitary}, and in a scan of $\kappa$~(Unitary) they capture most of the resonances where $\Delta(C^x_1)$ grows (Fig.~\ref{Fig:Resonances_Unitary_ECO}); a few additional narrow resonances may arise from higher-order processes. Away from the resonances, the spectral deviations are small and decrease with $L$, so the response requires no fine tuning of the scanned parameters.\\

%%%%%%%%%%%%%%%%%%%%%%%%%%%%%%%%%%%%%%%%%%%%%%%%%
\begin{figure}[tb]
\centering
\includegraphics[width=\columnwidth]{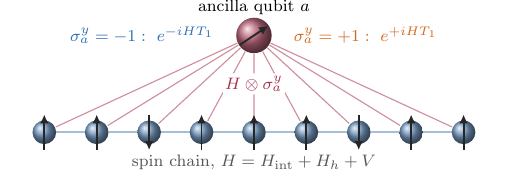}
\caption{\textbf{Single-ancilla embedding of the anti-unitary drive.} One ancilla qubit $a$ couples to every term of $H$ through $e^{iT_1H\otimes\sigma^y_a}$, so the chain evolves forward ($\sigma^y_a=-1$) or backward ($\sigma^y_a=+1$) in time. A kick $\mathcal{R}\otimes\sigma^z_a$ each period completes the unitary, $T_1$-periodic drive that realizes $A_F$.}
\label{Fig:ancilla}
\end{figure}
%%%%%%%%%%%%%%%%%%%%%%%%%%%%%%%%%%%%%%%%%%%%%%%%%
% \par\smallskip
\noindent
\textbf{\textit{A Realization of the Anti-Unitary Drive:}} Writing $\psi={\rm Re}\,\psi+i\,{\rm Im}\,\psi$ in the $\sigma^z$ basis, the map $|\psi\rangle\mapsto|\Psi\rangle={\rm Re}\,\psi\otimes|0\rangle_a+{\rm Im}\,\psi\otimes|1\rangle_a$~\cite{Casanova2011,Zhang2015_Embedding} sends $e^{-iHT_1}\mapsto e^{iT_1H\otimes\sigma^y_a}$ for the real $H$ of Eq.~\eqref{Eq:Hamiltonian_1_V4}, $\hat K\mapsto\openone\otimes\sigma^z_a$, and $\mathcal{R}\mapsto\mathcal{R}\otimes\openone$, while observables $O$ that are real in the $\sigma^z$ basis map to $O\otimes\openone$. Hence
\begin{equation}
\tilde U_F=(\mathcal{R}\otimes\sigma^z_a)\,e^{iT_1H\otimes\sigma^y_a}
\label{eq:embed}
\end{equation}
reproduces all their stroboscopic expectation values with the same period $T_1$. Because $\sigma^y_a$ is conserved during each interval, $e^{iT_1H\otimes\sigma^y_a}$ is forward or backward evolution under $H$, controlled by the ancilla, as in interferometric Loschmidt-echo and OTOC protocols; the $\sigma^z_a$ kick exchanges the two branches every period. The ancilla couples to every term of $H$, so the embedded drive is not a local chain. For real initial states and observables, the stroboscopic data of $A_F$ also coincide with those of the ancilla-free sequence $e^{-iHT_1}\to\mathcal{R}\to e^{+iHT_1}\to\mathcal{R}\to\cdots$, which reverses $H$ in every other cycle. That sequence has period $2T_1$. It probes Eq.~\eqref{eq:twocycle}, the conservation of $m^x$, and the resonances of Eq.~\eqref{Resonance_Anti-Unitary}, but the subharmonic character of the response refers to the anti-unitary $T_1$-periodicity, which becomes a unitary one only in the embedded system.\\

\noindent\textit{Conclusion:}
Inserting time reversal into the drive cycle turns dynamical freezing into a route to time-crystalline order in a clean, short-range, non-integrable chain. Anti-unitarity makes the order parameter itself the emergent conserved quantity and pins the subharmonic frequency through an exact quasi-energy pairing that survives errors in the flip angle, while DF keeps the effective Hamiltonian local in a strong-drive regime beyond the reach of high-frequency prethermalization. Infinite-chain simulations show no detectable decay over accessible times and rigidity under variation of $h_x$, complementing finite-system spectral evidence; infinite-time stability at $L=\infty$ remains a separate question. The anti-unitary cycle requires only one ancilla and controlled forward and backward evolution (Fig.~\ref{Fig:ancilla}), as available in embedding quantum simulators~\cite{Zhang2015_Embedding}.

\bibliography{DTCV1}

\clearpage
\onecolumngrid
\clearpage

% Fresh numbering for Supplemental Material.
\setcounter{section}{0}
\setcounter{subsection}{0}
\setcounter{subsubsection}{0}
\setcounter{equation}{0}
\setcounter{figure}{0}
\setcounter{table}{0}
\renewcommand{\theequation}{S\arabic{equation}}
\renewcommand{\thefigure}{S\arabic{figure}}
\renewcommand{\thetable}{S\arabic{table}}
% Independent supplementary section and subsection numbering.
% Keep the same REVTeX Roman/letter convention, with an S namespace.
\renewcommand{\thesection}{S\Roman{section}}
\renewcommand{\thesubsection}{\thesection.\Alph{subsection}}
\renewcommand{\thesubsubsection}{\thesubsection.\arabic{subsubsection}}
% Unique PDF link destinations after resetting counters.
\renewcommand{\theHequation}{SM.\arabic{equation}}
\renewcommand{\theHfigure}{SM.\arabic{figure}}
\renewcommand{\theHtable}{SM.\arabic{table}}
\renewcommand{\theHsection}{SM.\arabic{section}}
\renewcommand{\theHsubsection}{SM.\arabic{section}.\arabic{subsection}}

\begin{center}
{\large\bfseries Supplemental Material for:
Discrete Time Crystals from Anti-Unitary Floquet Driving and Dynamical Freezing}
% \vspace{1em}
% {\normalsize Sagnik Chaudhuri, Anirban Das, Asmi Haldar, and Arnab Das\par}
% \vspace{0.35em}
% {\small Indian Association for the Cultivation of Science, School of Physical Sciences,\\
% 2A \& 2B Raja S. C. Mullick Road, Kolkata 700032, West Bengal, India\par}
% {\small Univ Toulouse, CNRS, Laboratoire de Physique Th\\'eorique, Toulouse, France\par}
\end{center}
\vspace{1em}

\section{First-order FDPT admixture}
Over two cycles, $A_F^2$ is generated by $H_1=H_{\rm int}+H_h+V$ for $0<t<T_1$ and $H_2=-H_{\rm int}+H_h+V$ for $T_1<t<2T_1$ [Eq.~\eqref{eq:twocycle}]. Treating $V$ as the perturbation, the unperturbed Floquet states are $\sigma^x$-basis configurations $|n\rangle$ with $U_0(2T_1)|n\rangle=e^{-i\Theta_n}|n\rangle$, where $\Theta_n=2T_1E_h(n)$ because the interaction energy cancels between the two intervals. Writing $U(2T_1)=U_0(2T_1)[1-i\Gamma+O(\epsilon^2)]$ with $\Gamma=\int_0^{2T_1}U_0^\dagger(t)VU_0(t)\,dt$, the first-order Floquet state is $|n\rangle+\sum_{m\neq n}c_{mn}|m\rangle$, with
\begin{equation}
c_{mn}=\frac{-i\,\Gamma_{mn}}{e^{i(\Theta_m-\Theta_n)}-1}.
\end{equation}
For $|m\rangle$ obtained by flipping a single spin $\sigma_0$, $\langle m|V|n\rangle=-\epsilon$, and the phase accumulates at rate $Z+D$ in the first interval and $Z-D$ in the second, so
\begin{equation}
\Gamma_{mn}=\frac{i\epsilon\,(e^{i\theta}-1)}{Z-D}-\frac{2i\epsilon D\,\bigl(e^{i(Z+D)T_1}-1\bigr)}{Z^2-D^2},
\end{equation}
with $\theta=\Theta_m-\Theta_n=2ZT_1$, which gives Eq.~\eqref{eq:admix}. The leading part of $\Gamma_{mn}$, the amplitude generated within one period, vanishes identically at $\theta\in2\pi\mathbb{Z}$: the two intervals interfere destructively, and only the part proportional to $D/Z$ produces resonances. For $U_F$, $H_{1,2}=H_{\rm int}\pm(H_h+V)$, and the same calculation for the unitary-protocol gives Eq.~\eqref{eq:admix} with $\theta=2DT_1$.

\section{Anti-unitary dynamics for fully polarized state and high-temperature pure state}
\label{sm:polarized}

\subsection{\texorpdfstring
{Dynamics of $m^x$ from $|\uparrow\uparrow\uparrow\ldots\rangle_x$}
{Dynamics of the longitudinal magnetization from the all-up state}}

% In main Fig.~\msref{Fig:AU_dynamics}{1}, we study the anti-unitary
% dynamics starting from
% $|\uparrow\uparrow\downarrow\uparrow
%   \uparrow\uparrow\downarrow\uparrow\ldots\rangle_x$.
Here we show that a stable period-doubled response is also observed for the anti-unitary protocol when the system is initialized in the fully polarized state
$|\uparrow\uparrow\uparrow\ldots\rangle_x$.\\

\begin{figure}[!tp]
\centering
\includegraphics[width=0.48\linewidth]{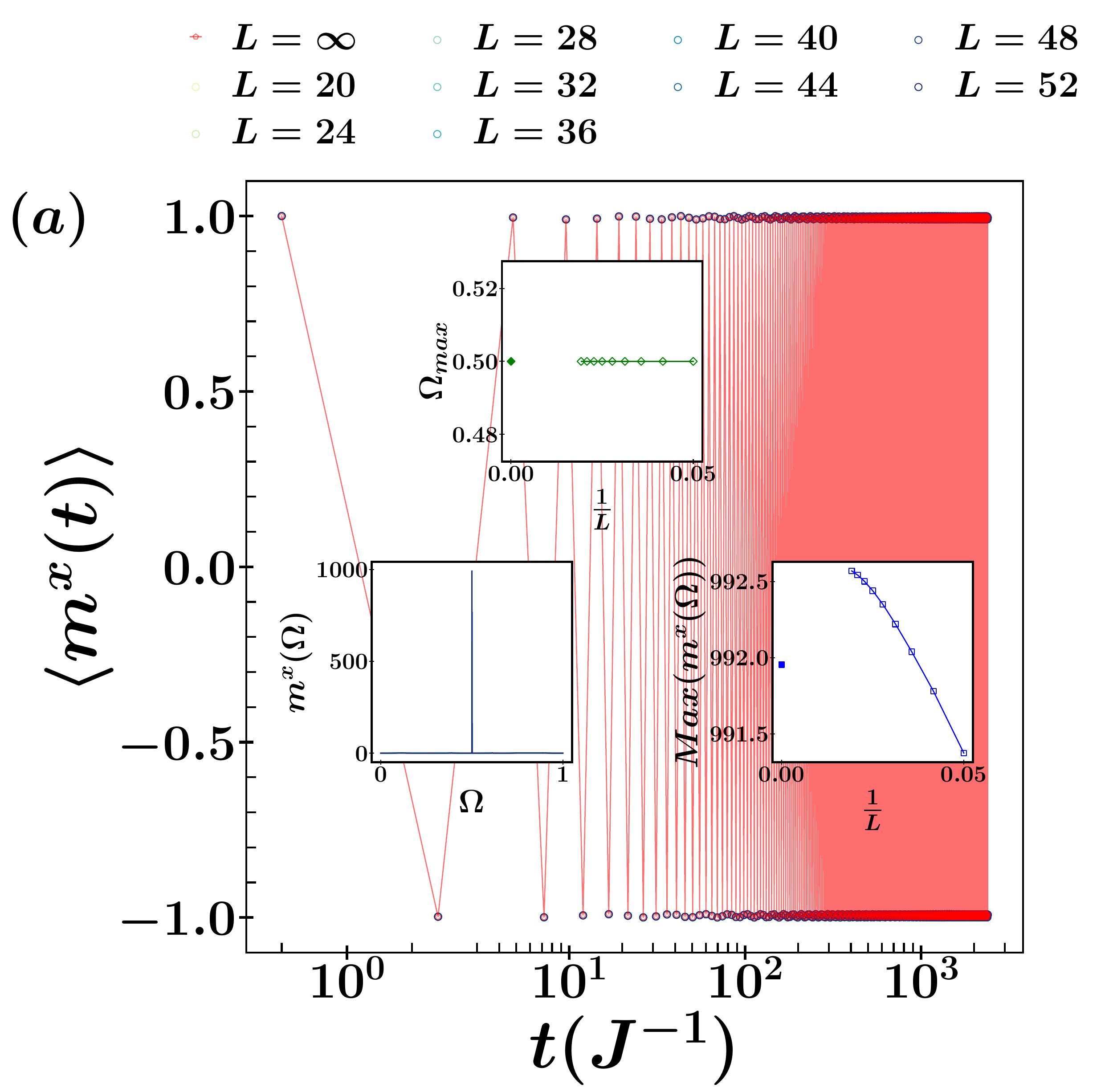}%
\hfill
\includegraphics[width=0.48\linewidth]{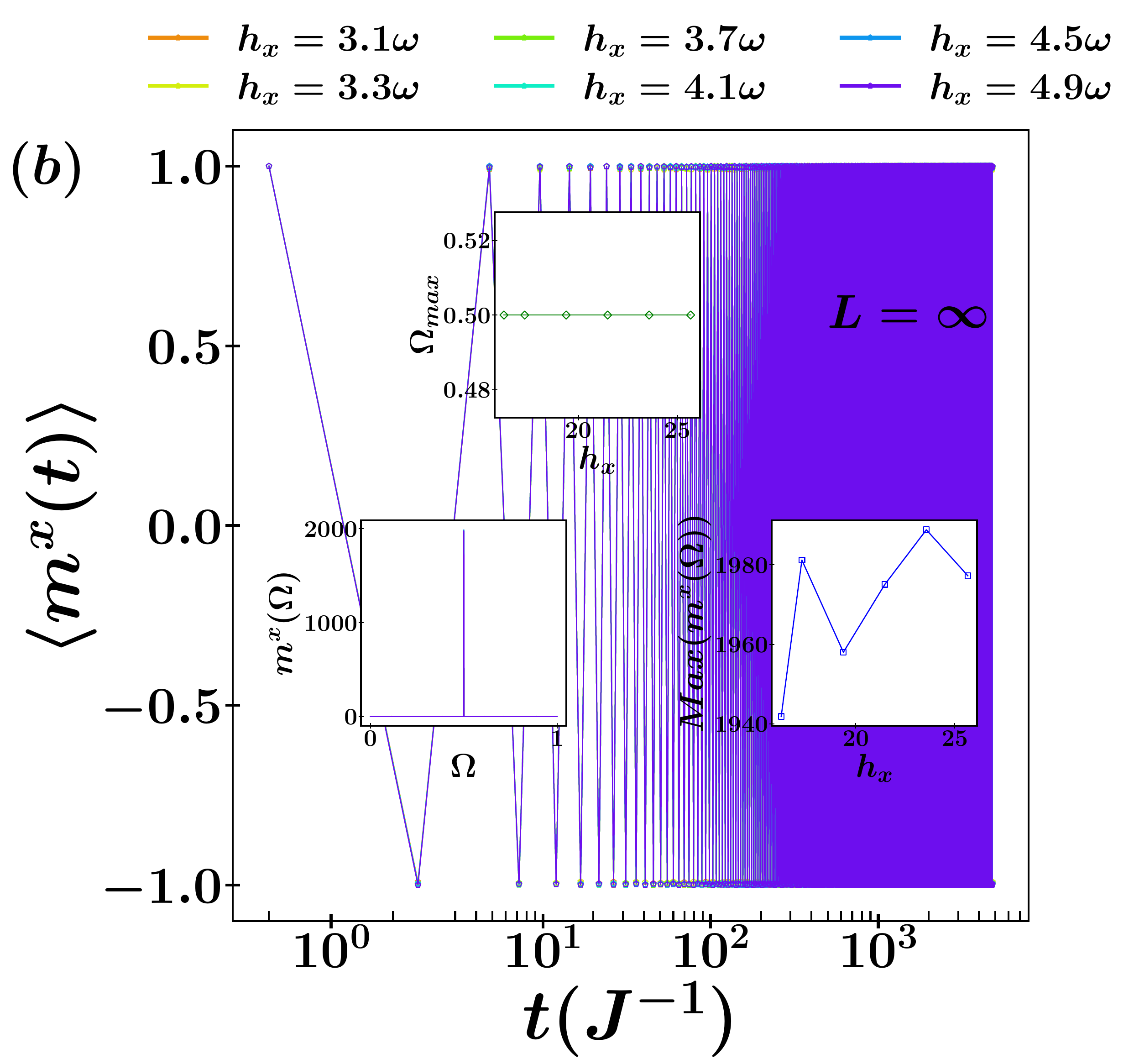}
\caption{\textbf{Period-doubled response of $m^x$ from the fully
polarized initial state under the anti-unitary protocol.}
\textbf{(a)} Stroboscopic dynamics of $\langle m^x(t)\rangle$
starting from $|\uparrow\uparrow\uparrow\ldots\rangle_x$ at
$h_x=3.3\omega$. We compare finite-chain TEBD results for the system sizes shown ($L=20$--$52$) with the iTEBD result at $L=\infty$. The Fourier analysis gives a dominant peak at
$\Omega=0.5$ for all sizes shown, with comparable peak strengths for the finite and infinite chains.
\textbf{(b)} Stroboscopic dynamics at $L=\infty$ for different values of $h_x$. The dominant Fourier peak remains at $\Omega=0.5$ throughout the field range shown, and its strength shows no systematic dependence on $h_x$. In both panels the rest of the Hamiltonian parameters were $J=2$, $\kappa=1.7$, $\epsilon=\sqrt{3}/1.5$, and
$\omega=2\pi/T_1=5\pi/3$.}
\label{sm:aupolar}
\end{figure}

As shown in Fig.~\ref{sm:aupolar}(a), the stroboscopic dynamics of $m^x$ exhibits period doubling for both finite chains and at $L=\infty$ over the time windows shown. The Fourier spectrum has a dominant peak at $\Omega=0.5$ for all the system sizes studied, including the infinite chain. The peak strength remains nonzero at $L=\infty$ and is comparable to the finite-chain values.\\

We further test the rigidity of the response by varying $h_x$ at $L=\infty$ while keeping the other parameters fixed. Figure~\ref{sm:aupolar}(b) shows that the dominant Fourier peak remains locked at $\Omega=0.5$ throughout the field range studied, while its strength shows no systematic dependence on $h_x$. The period-doubled response thus persists over the range of longitudinal fields shown.

\subsection{Dynamics for a high-temperature pure state}
\label{sm:superposition}

To probe the period-doubled response from a superposition of
$x$-basis configurations, we construct the initial state
\begin{equation}
\label{eq:sm-biased-pure-state}
\begin{gathered}
|\beta\rangle =
\frac{1}{\sqrt{\mathcal N_\beta}}
\sum_i c_i e^{\beta m_i^x}|i_x\rangle,\\
\mathcal N_\beta = \sum_i |c_i|^2 e^{2\beta m_i^x}.
\end{gathered}
\end{equation}
Here $|i_x\rangle$ denotes an $x$-basis configuration and
$m_i^x$ is its magnetization density. For each configuration, we draw the real and imaginary parts of $c_i=a_i+\mathrm{i}b_i$ independently and uniformly from $(-1,1)$, then multiply $c_i$ by $e^{\beta m_i^x}$ with $\beta=1.25$. The factor $\mathcal N_\beta^{-1/2}$ normalizes the state so that $\langle\beta|\beta\rangle=1$. The resulting state has nonzero initial magnetization and retains weight over a broad range of
$x$-basis configurations.

\begin{figure}[!tp]
\centering
\includegraphics[width=0.5\linewidth]{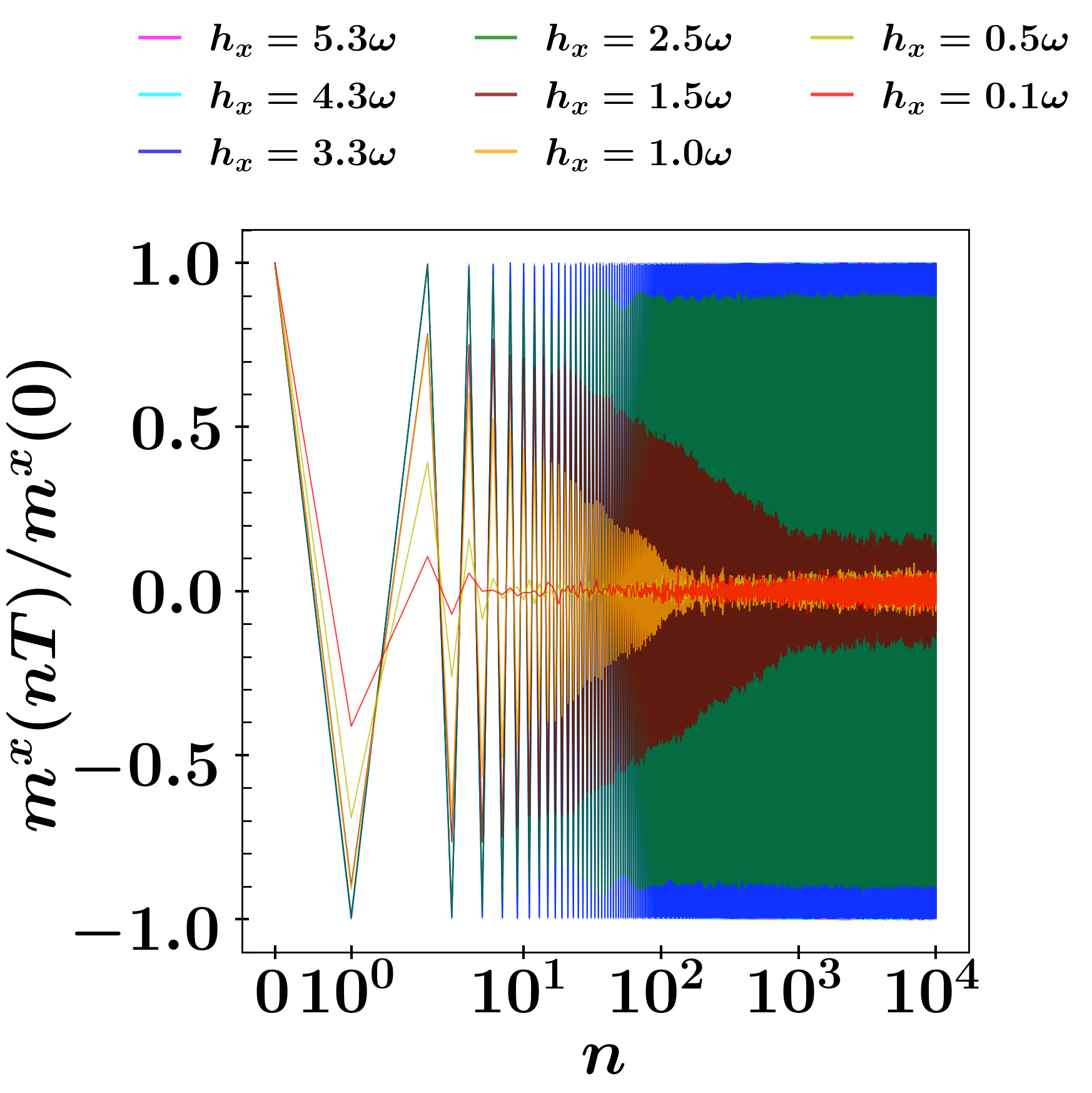}
\caption{\textbf{Period-doubled response of $m^x$ from a high-temperature pure state under the anti-unitary protocol.}
The stroboscopic magnetization, normalized by its initial value, $\langle m^x(nT)\rangle/\langle m^x(0)\rangle$, is plotted for the indicated values of $h_x$. Each cycle consists of evolution under $H$ for time $T_1$, followed by complex conjugation and a global $\pi$ rotation about the $y$ axis. The initial state is the normalized superposition $|\beta\rangle$ defined in Eq.~\eqref{eq:sm-biased-pure-state}, with $\beta=1.25$. For the stronger fields shown above approximately $3\omega$, no decay of the period-doubled oscillations is observed up to $10^4$ cycles.
The system size is $L=18$, and the remaining parameters are $J=2.0$, $\kappa=1.7$, $\omega=2\pi/T_1=5\pi/3$, and
$\epsilon=\sqrt{3}/1.5$.}
\label{sm:aubeta}
\end{figure}

We evolve this state under the anti-unitary protocol for different values of the longitudinal field $h_x$. As shown in Fig.~\ref{sm:aubeta}, the period-doubled oscillations of $m^x$ decay rapidly for the fields shown below $3\omega$. For the stronger fields shown above approximately $3\omega$, the period-doubled oscillations
show no observable decay throughout the simulation window of $10^4$ cycles in the $L=18$ chain.

\section{ECOs in the Unitary Protocol}
\label{sm:bond}

\subsection{Longitudinal-field dependence of ECOs}

In the unitary protocol, the global $\pi$ rotation reverses the longitudinal and transverse field terms while leaving the Ising interactions unchanged. The bond operators
\begin{equation}
\label{eq:sm-bond-operators}
C_r^x=\frac{1}{L}\sum_i \sigma_i^x\sigma_{i+r}^x,
\qquad r=1,2,
\end{equation}
are invariant under the rotation. To test whether these operators remain approximately conserved at finite $\epsilon$, we compare their expectation values in the Floquet eigenstates with their operator eigenvalues.

\begin{figure}[!tp]
\centering
\includegraphics[width=0.48\linewidth]{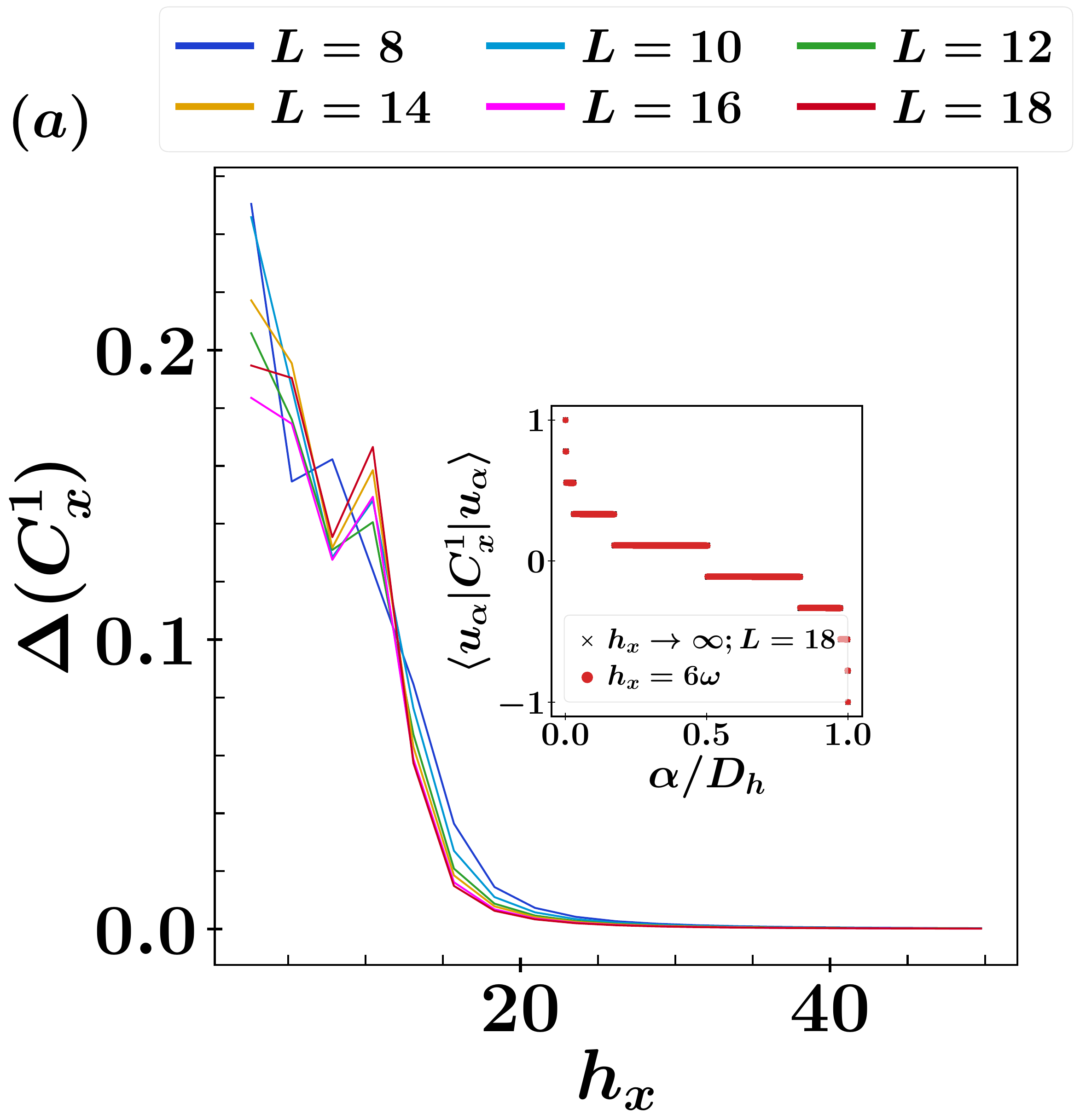}%
\hfill
\includegraphics[width=0.48\linewidth]{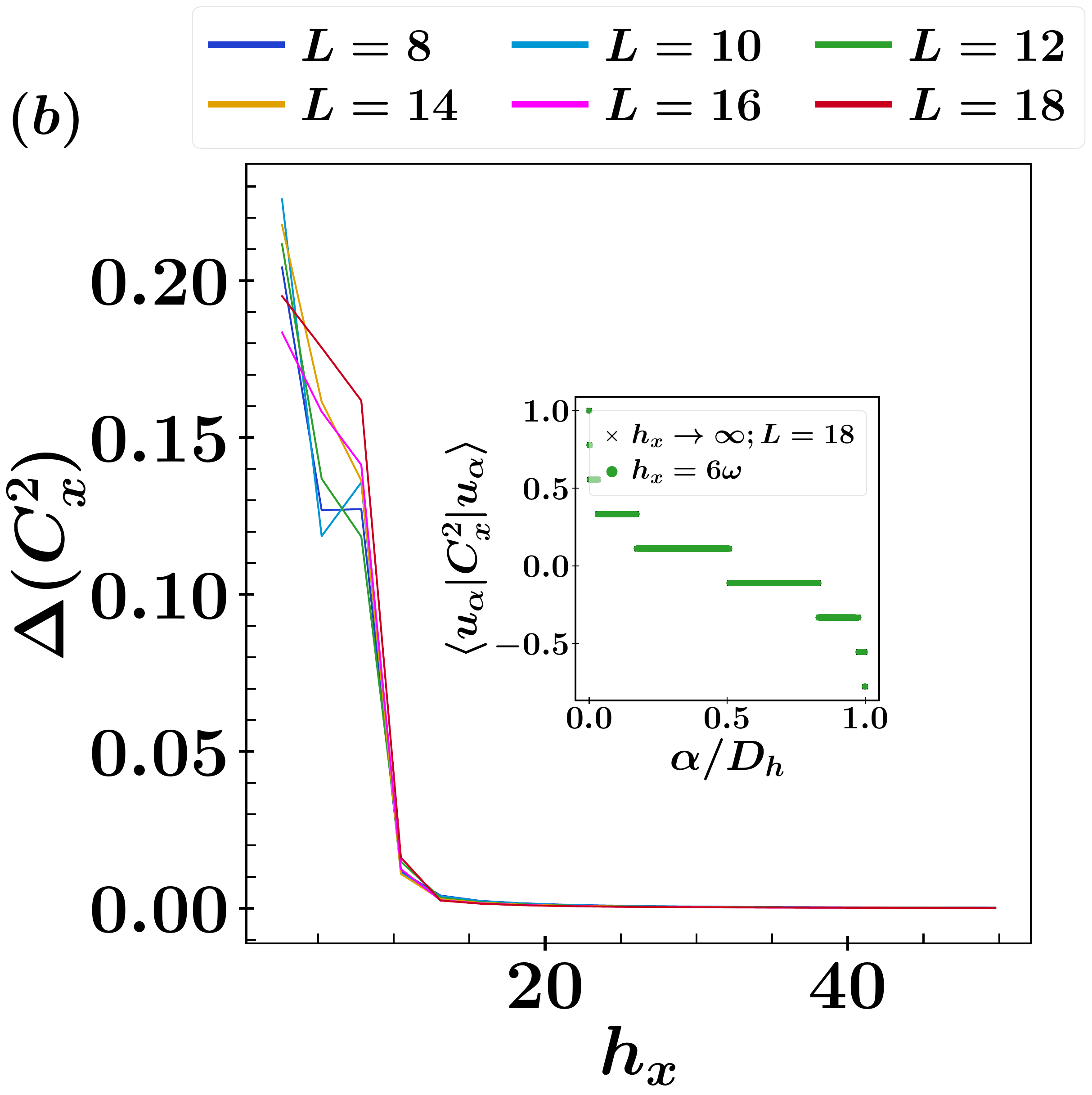}
% \caption{\textbf{$\Delta(C_r^x)$ as a function of the longitudinal field $h_x$ in the unitary protocol.}
% \textbf{(a)} Spectral deviation $\Delta(C_1^x)$, defined in
% Eq.~\eqref{eq:sm-bond-spectral-deviation}, as a function of the longitudinal field $h_x$ for $L=8$--$18$. The other parameters are $J=2$, $\kappa=1.7$,
% $\epsilon=\sqrt{3}/1.5$, and $\omega=2\pi/T_1=5\pi/3$. At sufficiently large $h_x$, the deviation is small and generally decreases with system size. The inset compares the ordered Floquet eigenstate expectation values of $C_1^x$ at $h_x=6\omega$ with the ordered eigenvalues of $C_1^x$. Both are plotted against $\alpha/D_H$, where $\alpha$ denotes the eigenvalue index in the respective sorted list.
% \textbf{(b)} The corresponding longitudinal-field scan for $C_2^x$, with the same parameters as in (a). The inset shows the analogous comparison for $C_2^x$ at $h_x=6\omega$. For both correlators, the ordered expectation values closely follow the steps in the
% operator spectra.}
\caption{\textbf{Longitudinal-field dependence of $\Delta(C_r^x)$ in the unitary protocol.}
\textbf{(a)} Spectral deviation $\Delta(C_1^x)$, defined in
Eq.~\eqref{eq:sm-bond-spectral-deviation}, as a function of $h_x$ for $L=8$--$18$.
\textbf{(b)} The corresponding scan for $\Delta(C_2^x)$.
In both panels, $J=2$, $\kappa=1.7$, $\epsilon=\sqrt{3}/1.5$,
and $\omega=2\pi/T_1=5\pi/3$. At sufficiently large $h_x$, the deviations are small and generally decrease with system size. \textbf{Insets:} At $h_x=6\omega$, we compare the sorted Floquet eigenstate expectation values
$\langle\mu_\alpha|\mathcal O|\mu_\alpha\rangle$ with the independently sorted eigenvalues of $\mathcal O$, where $\mathcal O=C_1^x$ in (a) and $\mathcal O=C_2^x$ in (b). Both sets of values are plotted against $\alpha/D_H$, where $\alpha$ is the index in the respective sorted list. The two sets agree closely in both panels.}
\label{sm:field}
\end{figure}

For $\mathcal O=C_1^x$ or $C_2^x$, let $o_n$ be the $n$th value in the sorted list of Floquet eigenstate expectation values and $\lambda_n(\mathcal O)$ the $n$th value in the independently sorted list of operator eigenvalues. We use the spectral deviation defined
in the main manuscript,  
\begin{equation}
\label{eq:sm-bond-spectral-deviation}
\Delta(\mathcal O)=\frac{1}{D_H}\sum_{n=1}^{D_H}
\bigl|o_n-\lambda_n(\mathcal O)\bigr|,
\end{equation}
where $D_H$ is the Hilbert-space dimension. Figure~\ref{sm:field} shows that the spectral deviations for both the bond correlators become small as $h_x$ increases and generally decrease with $L$ in the strong-field regime shown. At $h_x=6\omega$,
the ordered expectation values closely follow the steps in the corresponding operator spectra. This agreement, together with the dynamics reported in the Sec.~\ref{sm:patterns}, supports the identification of $C_1^x$ and $C_2^x$ as emergent approximately conserved operators in the unitary protocol beyond a certain value of $h_x$.

\subsection{Frequency dependence of bond ECO stability}
\begin{figure}[!tp]
\centering
\includegraphics[width=0.48\linewidth]{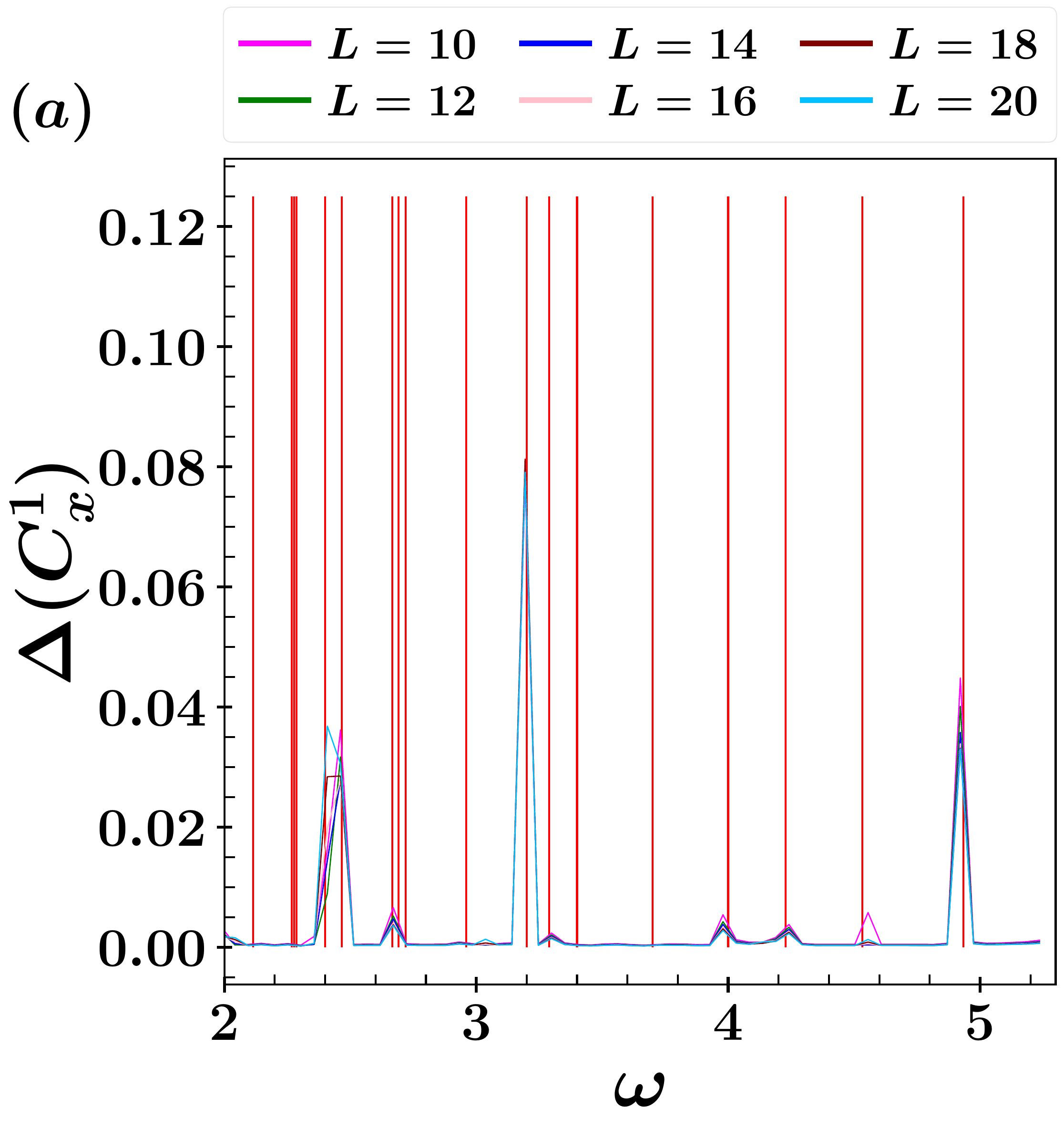}%
\hfill
\includegraphics[width=0.48\linewidth]{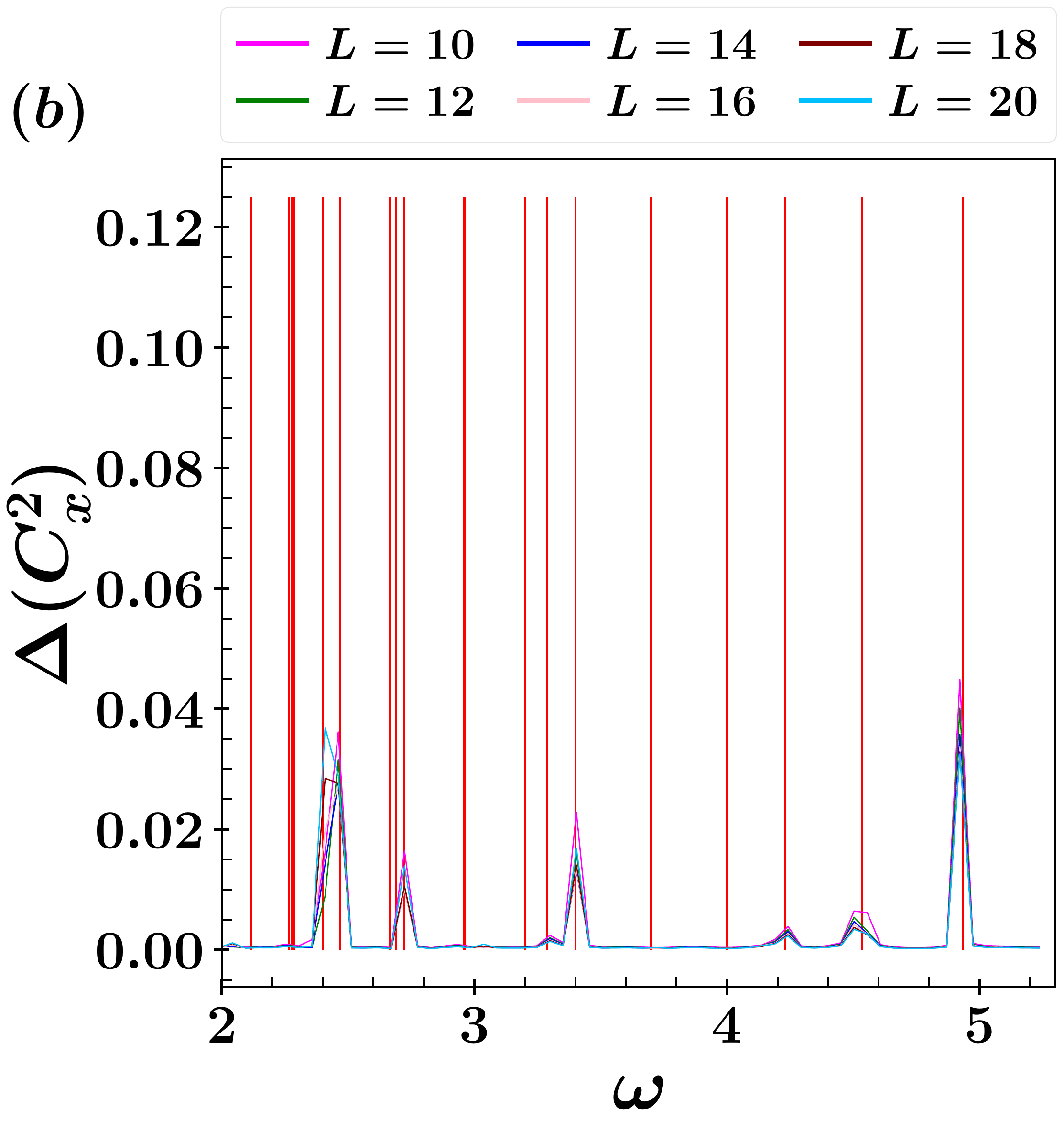}
\caption{\textbf{Frequency dependence of $\Delta(C_r^x)$ in the unitary protocol.}
\textbf{(a)} Spectral deviation $\Delta(C_1^x)$ and
\textbf{(b)} spectral deviation $\Delta(C_2^x)$ as functions of the drive frequency $\omega=2\pi/T_1$ for $L=10,12,14,16,18,20$. In both panels, $J=2$, $\kappa=1.7$, $h_x=10\pi$, and $\epsilon=\sqrt{3}/1.5$ are held fixed. The red vertical lines mark the frequencies obtained from the first-order FDPT denominator condition, Eq.~\eqref{eq:sm-unitary-resonance}. At these few isolated resonance points the quality of the conservation deteriorate.}
\label{sm:resonances}
\end{figure}

% In main Fig.~\msref{Fig:Resonances_Unitary_ECO}{4}, we vary the
% next-nearest-neighbor coupling $\kappa$ at fixed frequency.
Here we examine how the approximate conservation of $C_1^x$ and $C_2^x$ depends on the drive frequency. We plot $\Delta(C_1^x)$ and $\Delta(C_2^x)$ as functions of $\omega=2\pi/T_1$ for different system sizes $L$, keeping $J$, $\kappa$, $h_x$, and $\epsilon$ fixed~(we have kept rest of the parameters same as before and fixed $h_x=10\pi$). For two unperturbed configurations that differ by a single spin flip denoted by $\sigma_0$, the first-order Floquet--Dyson perturbation theory (FDPT) denominator vanishes when
\begin{equation}
\label{eq:sm-unitary-resonance}
\sigma_0\bigl[J(\sigma_1+\sigma_{-1})
-\kappa(\sigma_2+\sigma_{-2})\bigr]
=\frac{p\omega}{4}, \qquad p\in\mathbb{Z},
\end{equation}
Here $\sigma_j=\pm1$. The vertical red lines in
Fig.~\ref{sm:resonances} mark the resonance frequencies predicted by leading-order FDPT. Figure~\ref{sm:resonances} shows that the spectral deviations of both the bond correlators remain small over broad frequency intervals. Near a few marked frequencies which corresponds to the resonance condition predicted by FDPT, the deviation increases and its dependence on $L$ becomes less systematic,    

\section{Unitary dynamics for different initial states and observables}
\label{sm:patterns}

In this section, we present the unitary dynamics from three
$x$-basis initial states: the fully polarized state
$|\uparrow\uparrow\uparrow\ldots\rangle_x$, the N\'eel state
$|\uparrow\downarrow\uparrow\downarrow\ldots\rangle_x$, and the
four-site pattern $|\uparrow\uparrow\downarrow\downarrow\ldots\rangle_x$.
We monitor the uniform magnetization $m^x$ for the fully polarized
state. For the other two states, the uniform magnetization vanishes
in both the initial pattern and its global spin-flip partner.
We therefore monitor the staggered and four-site magnetizations,
respectively, which change sign under the global spin flip.

\subsection{Unitary dynamics from the fully polarized state}
\label{sm:unitary-polarized}

\begin{figure}[!tp]
\centering
\includegraphics[width=0.48\linewidth]{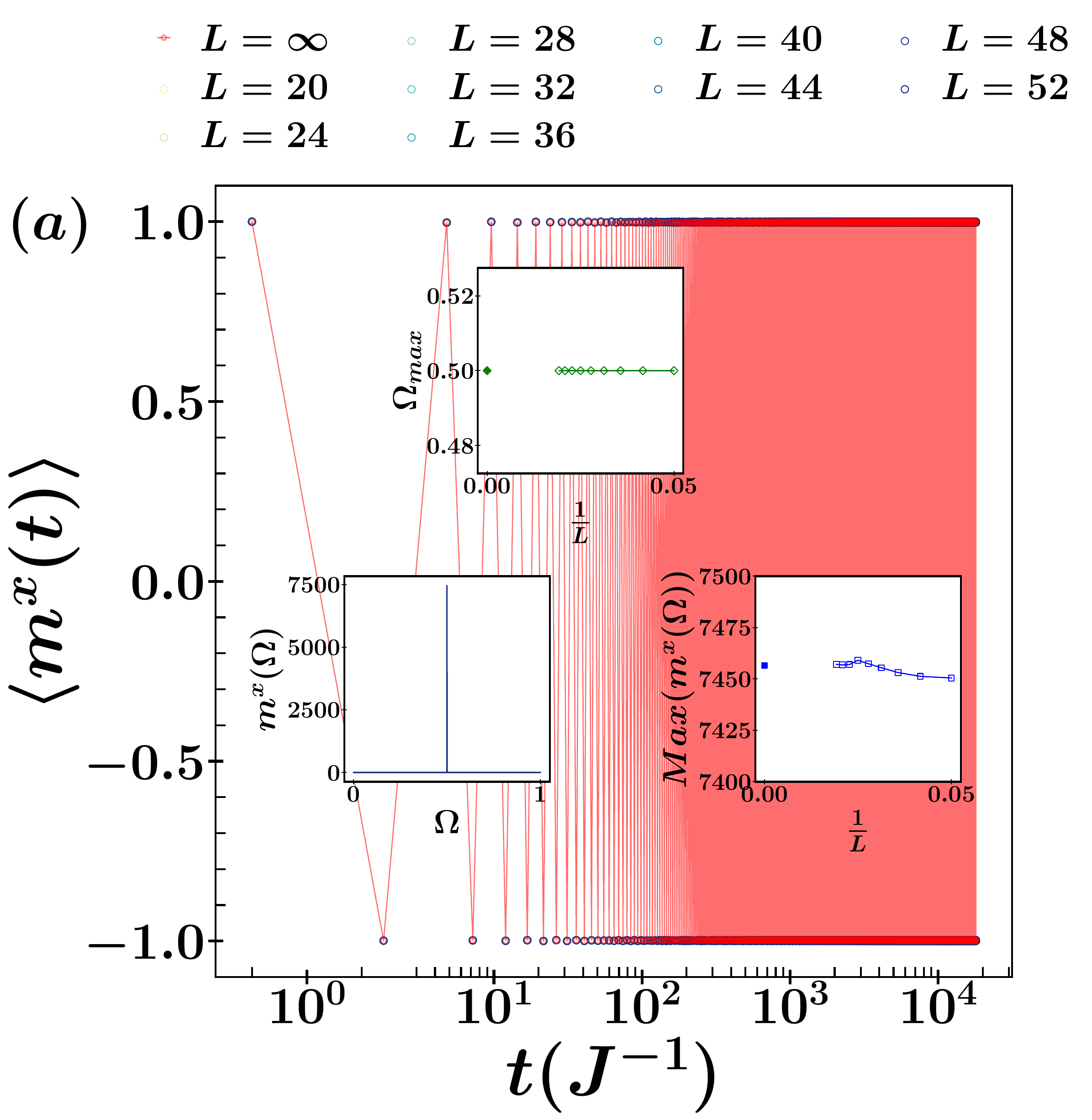}\hfill
\includegraphics[width=0.48\linewidth]{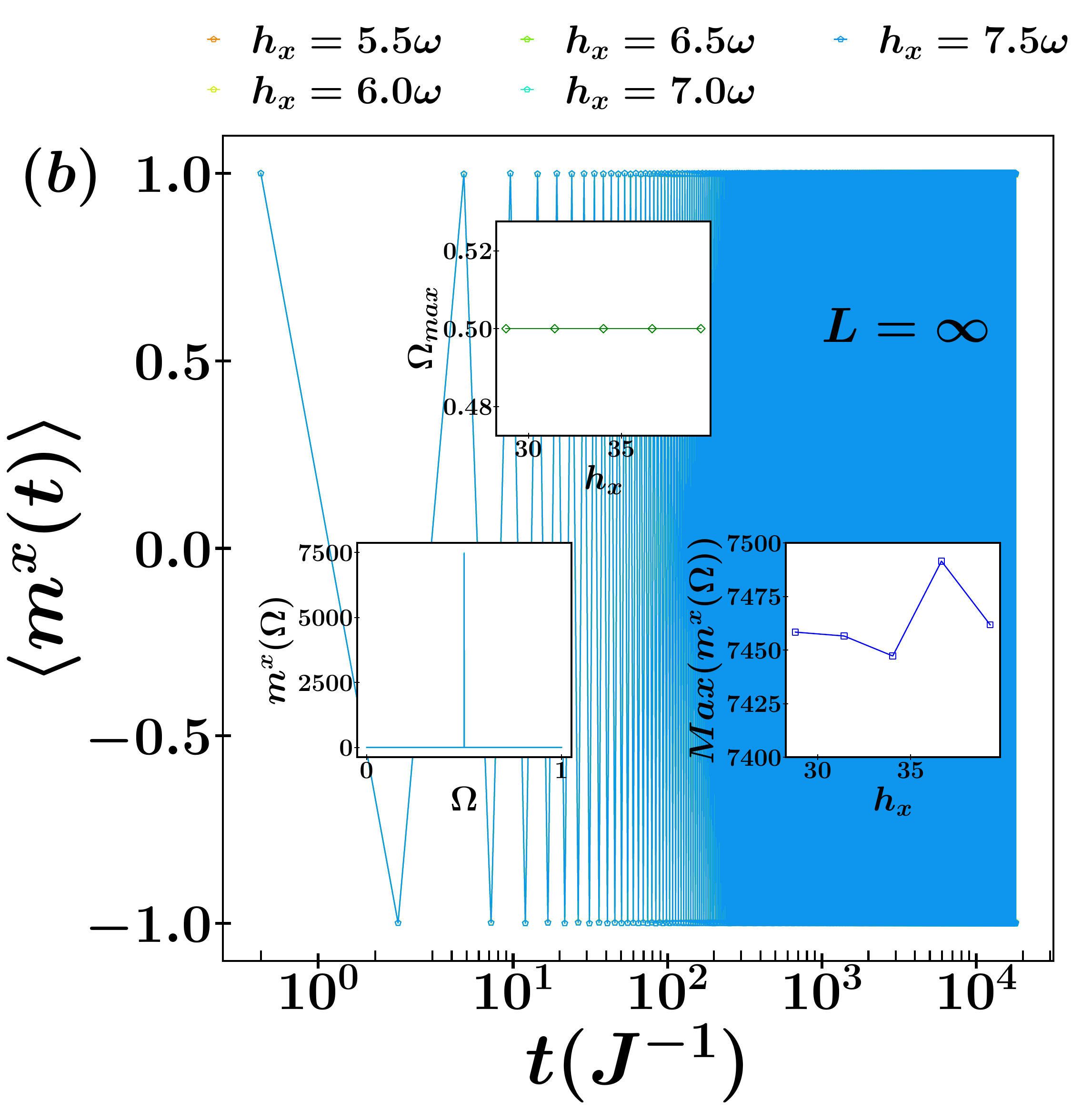}
\caption{\textbf{Period-doubled magnetization from the fully polarized
state under the unitary protocol.}
\textbf{(a)} Stroboscopic magnetization $\langle m^x(t)\rangle$ under $U_F$ from $|\uparrow\uparrow\uparrow\cdots\rangle_x$ at $h_x=6\omega$. We compare finite-chain TEBD results with the
infinite-chain iTEBD result. No decay of the period-doubled
response is detectable up to times of order $10^4J^{-1}$.
The bottom-left inset shows the Fourier spectrum. The top inset shows the position $\Omega_{\max}$ of the dominant Fourier peak versus $1/L$, with $L=\infty$ shown as a solid marker. The peak is at $\Omega=0.5$ for all sizes shown. The bottom-right inset shows the peak strength at $\Omega=0.5$ versus $L$. \textbf{(b)} Stroboscopic magnetization at $L=\infty$ for several values of $h_x$. The bottom-left inset shows the Fourier spectra,
and the top and bottom-right insets show the dominant peak
position and the peak strength at $\Omega=0.5$ versus $h_x$,
respectively. The peak remains locked at $\Omega=0.5$, with no systematic drift in its strength across the field range shown. In both panels, $J=2$, $\kappa=1.7$,
$\omega=2\pi/T_1=5\pi/3$, and $\epsilon=\sqrt{3}/1.5$.}
\label{Fig:U_dynamics}
\end{figure}

We initialize the chain in the fully polarized state
$|\uparrow\uparrow\uparrow\cdots\rangle_x$ and track the
stroboscopic magnetization $m^x$ under the unitary protocol.
Figure~\ref{Fig:U_dynamics}(a) shows a period-doubled response in both finite chains and the infinite chain, with the dominant Fourier peak locked at $\Omega=0.5$. To test the rigidity of the response, we vary $h_x$ at $L=\infty$. The peak remains at $\Omega=0.5$, and its strength shows no systematic drift across the fields studied [Fig.~\ref{Fig:U_dynamics}(b)].

\subsection{\texorpdfstring
  {Dynamics from $|\uparrow\downarrow\uparrow\downarrow\ldots\rangle_x$
   using $m^x_{\uparrow\downarrow}$}
  {Dynamics from the Neel state using the staggered magnetization}}

\begin{figure}[!tp]
\centering
\includegraphics[width=0.48\linewidth]{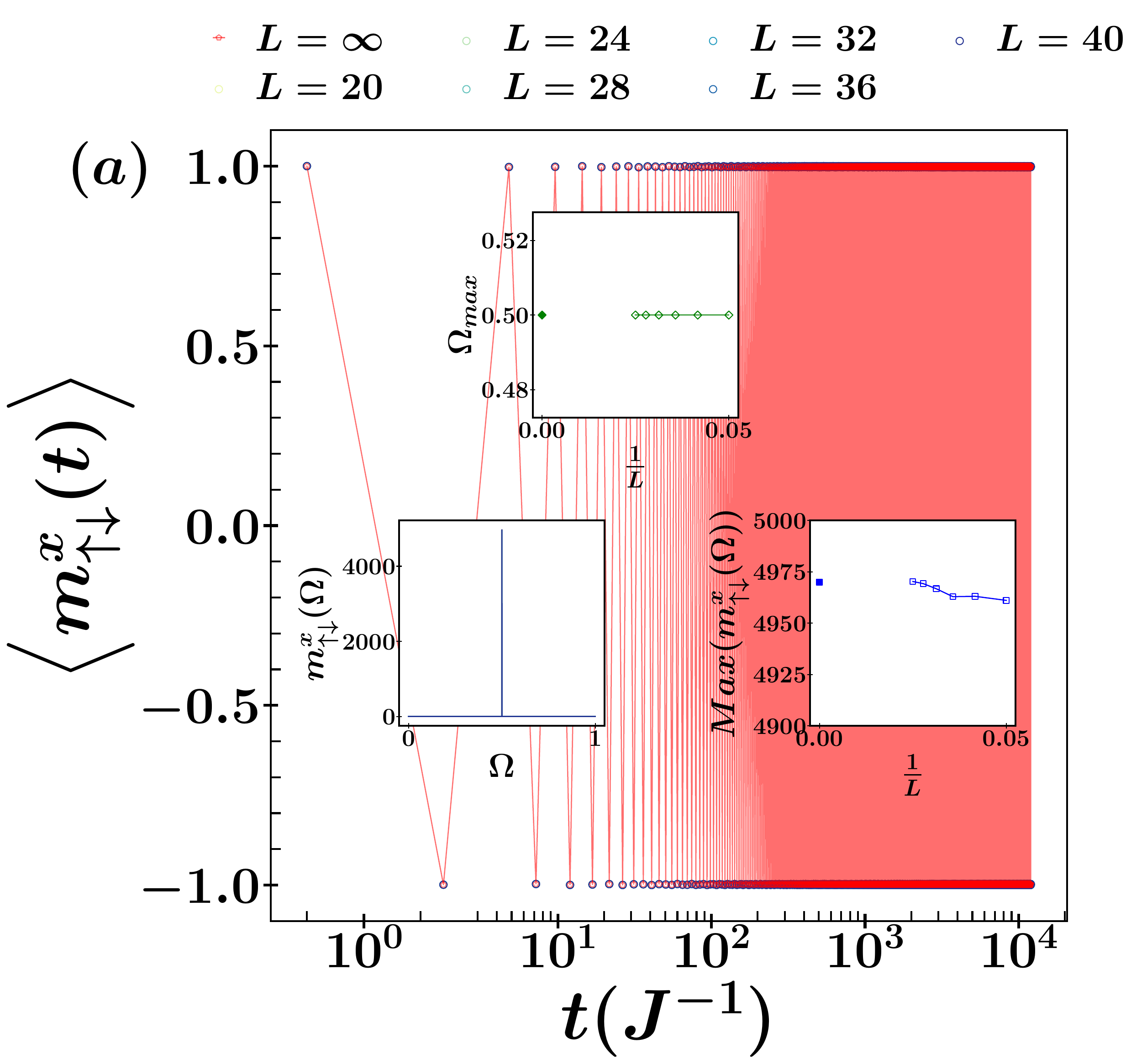}%
\hfill
\includegraphics[width=0.48\linewidth]{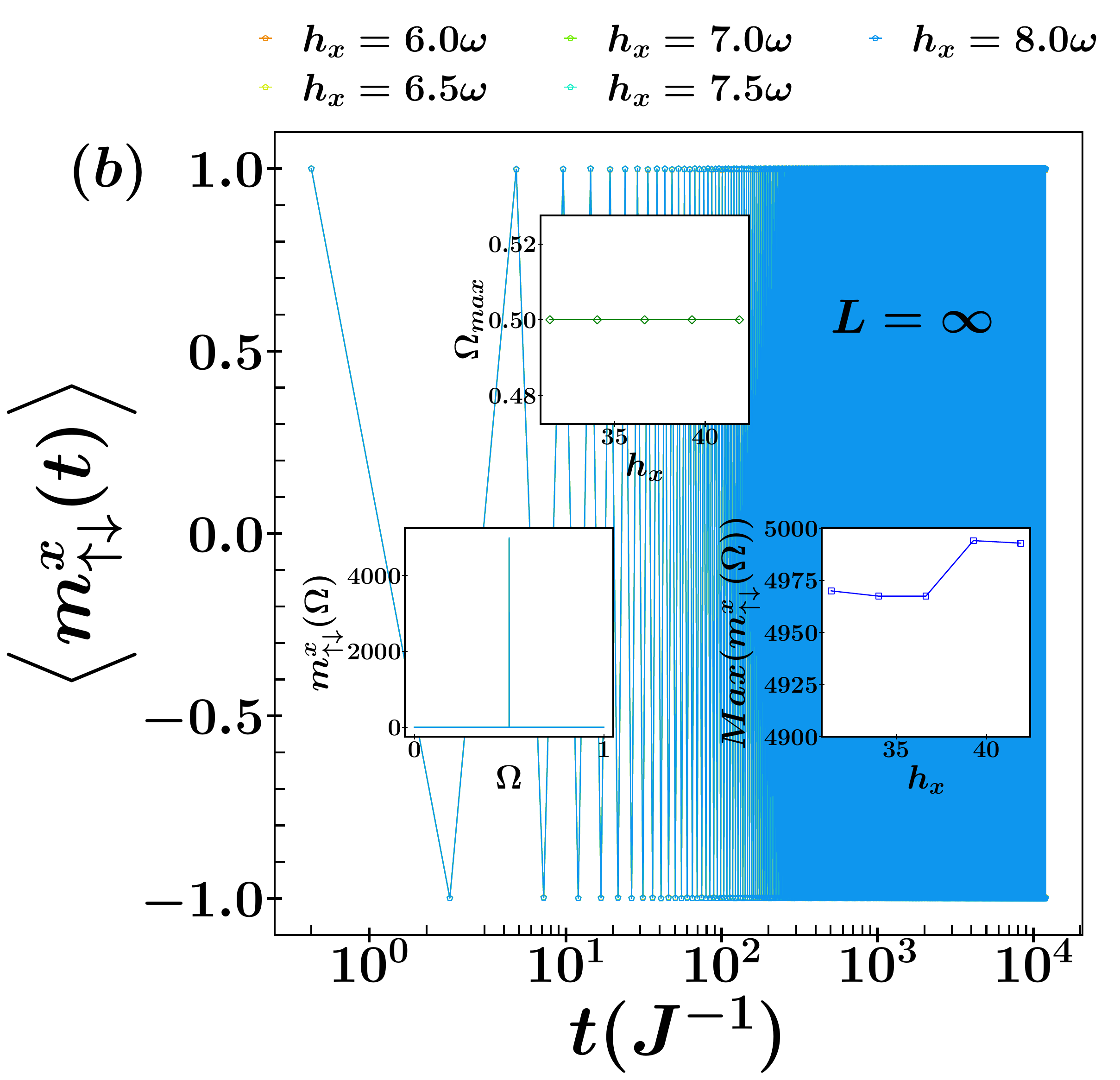}
\caption{\textbf{Period-doubled response of
$m^x_{\uparrow\downarrow}$ from the N\'eel initial state.}
\textbf{(a)} Stroboscopic dynamics of
$\langle m^x_{\uparrow\downarrow}(t)\rangle$, defined in
Eq.~\eqref{eq:sm-neel-magnetization}, starting from
$|\uparrow\downarrow\uparrow\downarrow\ldots\rangle_x$ for
$L=20,24,\ldots,40$ and $L=\infty$. The parameters are $J=2$,
$\kappa=1.7$, $\omega=2\pi/T_1=5\pi/3$, $\epsilon=\sqrt{3}/1.5$,
and $h_x=6\omega$. The bottom-left inset shows the discrete Fourier
spectrum as a function of the normalized frequency $\Omega$.
The top inset shows the position $\Omega_{\max}$ of the dominant
Fourier peak as a function of $1/L$. The peak remains at
$\Omega=0.5$ for the finite chains and at $L=\infty$ (solid marker).
The bottom-right inset shows the strength of the peak at
$\Omega=0.5$, which has no systematic dependence on $L$ over the
sizes shown.
\textbf{(b)} Stroboscopic dynamics at $L=\infty$ for different
values of $h_x$, with the other parameters fixed as in (a).
The bottom-left inset shows the Fourier spectra, the top inset
shows $\Omega_{\max}$ as a function of $h_x$, and the bottom-right
inset shows the corresponding peak strength. The peak remains at
$\Omega=0.5$ and its strength shows no systematic trend over the
field range studied.}
\label{sm:neel}
\end{figure}

We initialize the system in the N\'eel state
$|\uparrow\downarrow\uparrow\downarrow\ldots\rangle_x$ and track the
stroboscopic dynamics of the staggered magnetization in the
$x$ direction,
\begin{equation}
 m^x_{\uparrow\downarrow}
 = \frac{1}{L}\sum_{j=1}^{L}(-1)^{j-1}\sigma_j^x,
 \label{eq:sm-neel-magnetization}
\end{equation}
Since $m^x$, $C_1^x$, and $C_2^x$ do not distinguish this state from its spin-flipped partner, we use the staggered magnetization in Eq.~\eqref{eq:sm-neel-magnetization}, which takes opposite values in the two states. We perform the same analysis as for the fully polarized initial
state and find a stable period-doubled response over the times studied. In Fig.~\ref{sm:neel}(a), the dominant Fourier peak remains locked at $\Omega=0.5$ for both finite and infinite chains. In Fig.~\ref{sm:neel}(b), we vary $h_x$ from $6\omega$ to $8\omega$ at $L=\infty$ and find that the peak remains at $\Omega=0.5$, showing that the subharmonic response is rigid over this range.

\subsection{\texorpdfstring
  {Dynamics from
   $|\uparrow\uparrow\downarrow\downarrow\ldots\rangle_x$
   using $m^x_{\uparrow\uparrow\downarrow\downarrow}$}
  {Dynamics from the up-up-down-down spin pattern}}

\begin{figure}[!tp]
\centering
\includegraphics[width=0.48\linewidth]{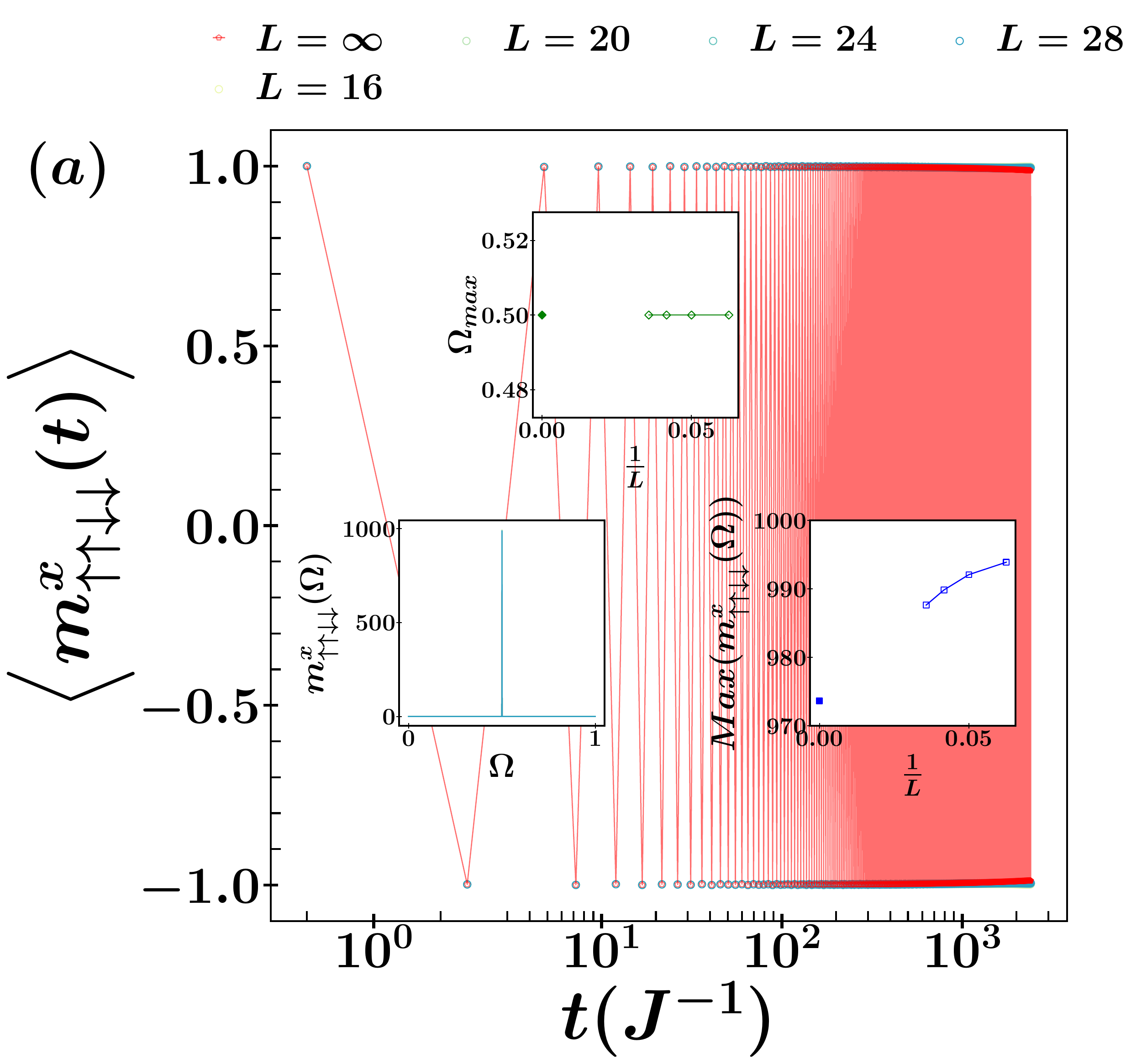}%
\hfill
\includegraphics[width=0.48\linewidth]{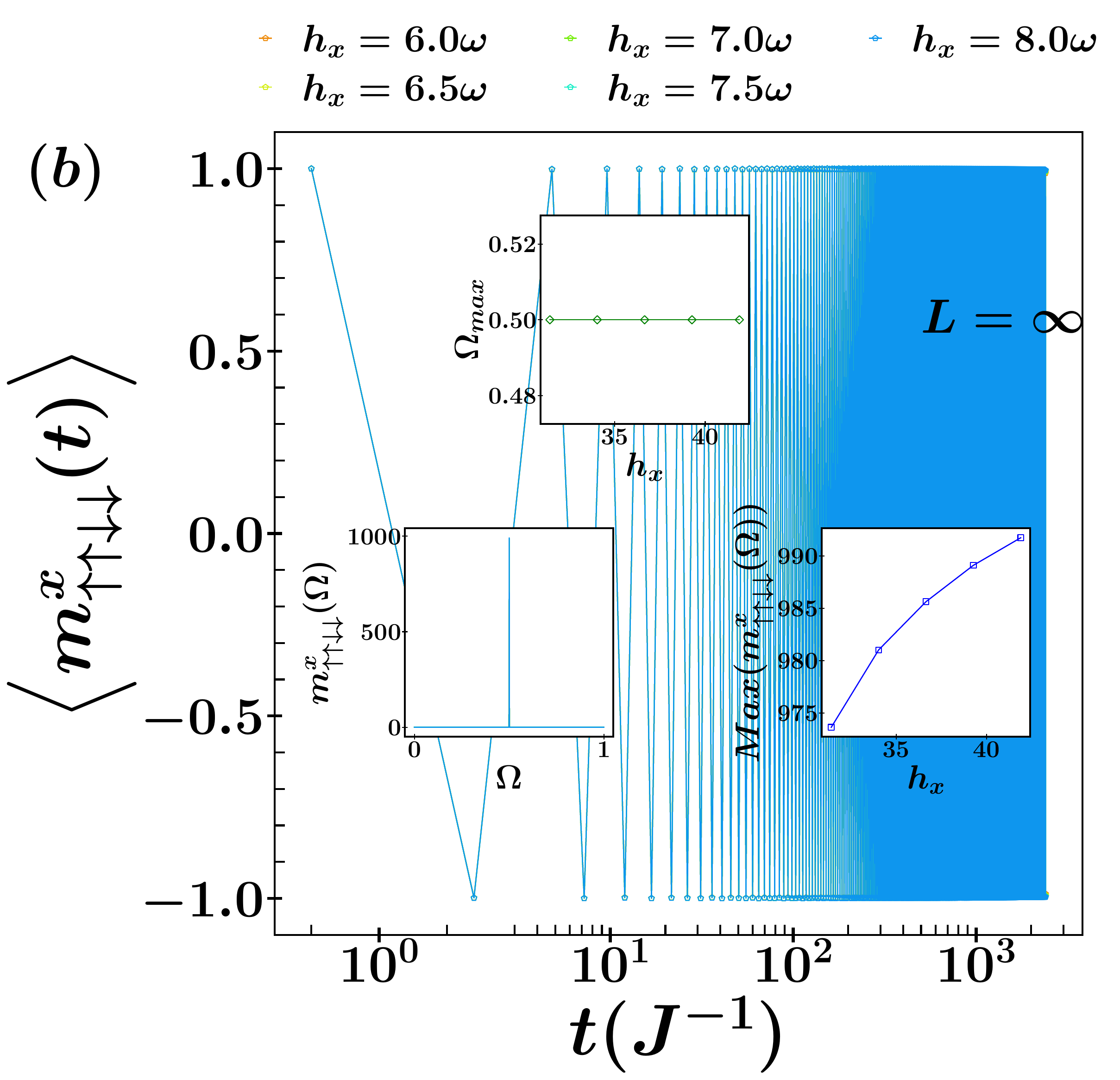}
\caption{\textbf{Period-doubled response of
$m^x_{\uparrow\uparrow\downarrow\downarrow}$ from the initial state
$|\uparrow\uparrow\downarrow\downarrow\ldots\rangle_x$.}
\textbf{(a)} Stroboscopic dynamics of
$\langle m^x_{\uparrow\uparrow\downarrow\downarrow}(t)\rangle$,
with the observable defined in Eq.~\eqref{eq:sm-uudd-magnetization},
for different system sizes and at $L=\infty$. The parameters are
$J=2$, $\kappa=1.7$, $\omega=2\pi/T_1=5\pi/3$,
$\epsilon=\sqrt{3}/1.5$, and $h_x=6\omega$. The bottom-left inset
shows the discrete Fourier spectrum of the stroboscopic signal.
The top inset shows the position $\Omega_{\max}$ of the dominant
Fourier peak as a function of $1/L$. The peak is at $\Omega=0.5$
for the finite chains and at $L=\infty$ (solid marker). The
bottom-right inset shows the strength of the peak at $\Omega=0.5$
as a function of $L$.
\textbf{(b)} Stroboscopic dynamics at $L=\infty$ for different
values of $h_x$, with the other parameters fixed as in (a).
The bottom-left inset shows the Fourier spectra. The top inset
shows that $\Omega_{\max}$ remains at $0.5$ as $h_x$ is varied,
and the bottom-right inset shows the corresponding peak strength
as a function of $h_x$.}
\label{sm:uudd}
\end{figure}

We next initialize the system in the state
$|\uparrow\uparrow\downarrow\downarrow\ldots\rangle_x$ and track
the stroboscopic dynamics of the operator
\begin{equation}
 m^x_{\uparrow\uparrow\downarrow\downarrow}
 = \frac{1}{L}\sum_{j=1,5,9,\ldots}
 \bigl(\sigma_j^x+\sigma_{j+1}^x
       -\sigma_{j+2}^x-\sigma_{j+3}^x\bigr).
 \label{eq:sm-uudd-magnetization}
\end{equation}
The operator in Eq.~\eqref{eq:sm-uudd-magnetization} takes the values $+1$ and $-1$ in the initial state and its globally flipped partner, respectively, and is zero in the other two other translational partners of the $x$ bit-string state. At $h_x=6\omega$, with the remaining parameters unchanged ($J=2$, $\kappa=1.7$, $\omega=5\pi/3$, and $\epsilon=\sqrt{3}/1.5$), we observe stable period doubling in both finite and infinite chains over the times studied. Figure~\ref{sm:uudd}(a) shows that the dominant Fourier peak remains locked at $\Omega=0.5$, including at $L=\infty$. In Fig.~\ref{sm:uudd}(b), we vary $\epsilon/h_x$ by changing $h_x$ at fixed $\epsilon$. The peak remains at $\Omega=0.5$ over the range studied, showing the rigidity of the subharmonic response.

\section{Response of $m^x$ for different $x$-basis product states in the unitary protocol}
\label{sm:generic}

\begin{figure}[!tp]
\centering
\begin{minipage}[t]{0.45\linewidth}
\noindent\makebox[\linewidth][c]{\textbf{(a)}}\par\nobreak
\vspace{2pt}
\noindent\includegraphics[width=\linewidth]{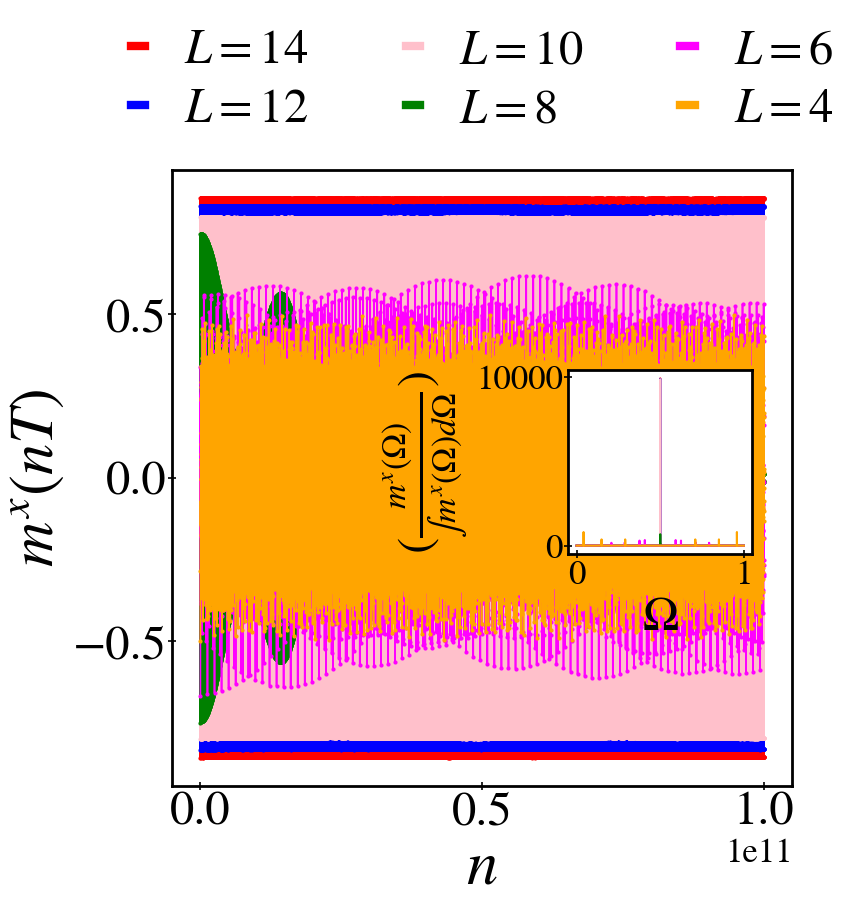}
\end{minipage}\hfill%
\begin{minipage}[t]{0.45\linewidth}
\noindent\makebox[\linewidth][c]{\textbf{(b)}}\par\nobreak
\vspace{2pt}
\noindent\includegraphics[width=\linewidth]{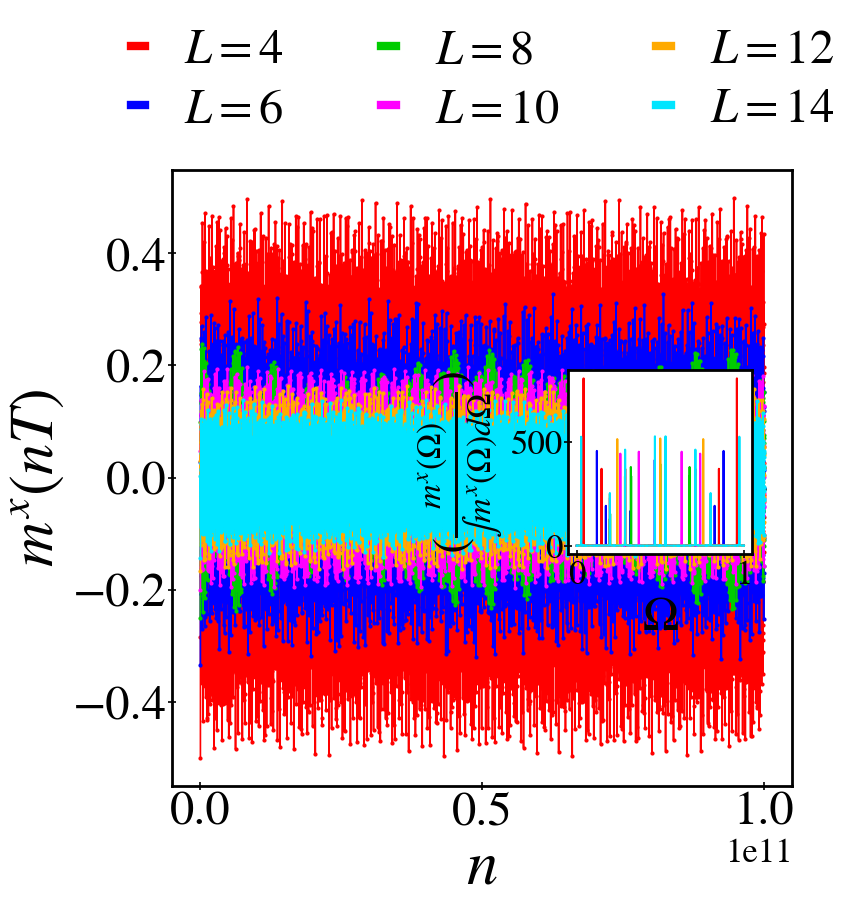}
\end{minipage}

\par\medskip
\begin{minipage}[t]{0.45\linewidth}
\noindent\makebox[\linewidth][c]{\textbf{(c)}}\par\nobreak
\vspace{2pt}
\noindent\includegraphics[width=\linewidth]{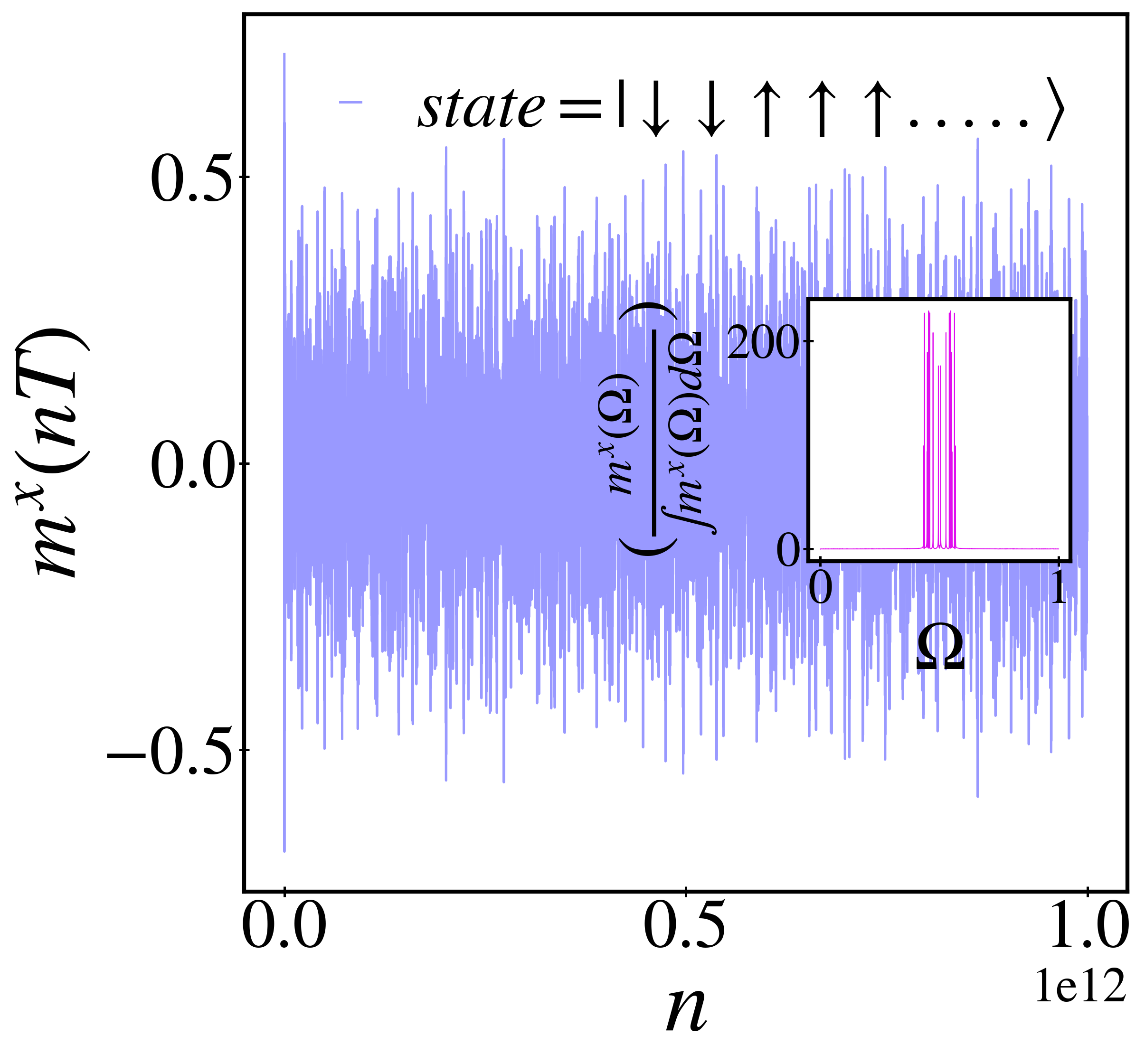}
\end{minipage}

\caption{\textbf{Response of $m^x$ for generic product states in the
unitary protocol.}
\textbf{(a)} Stroboscopic dynamics of $m^x$ starting from a single
spin flip in a fully polarized $x$-basis state, for different
system sizes $L$. The inset shows the Fourier spectrum. The
period-doubled response becomes more stable as $L$ increases.
\textbf{(b)} Stroboscopic dynamics of $m^x$ starting from a
N\'eel state with a single spin flipped, for different system
sizes $L$. Within the bond ECO sector, configurations with
opposite magnetizations can be connected by two spin flips.
The Fourier spectrum in the inset shows additional components
alongside the dominant subharmonic response at $\Omega=0.5$.
\textbf{(c)} Long-time dynamics of $m^x$ starting from two adjacent
spin flips in a fully polarized $x$-basis state, for $L=14$.
In this sector, the bond ECOs do not fix $|m^x|$, and the inset
shows additional frequency components alongside the subharmonic
response at $\Omega=0.5$.
In all panels, $J=2.0$, $\kappa=1.7$, $\omega=2\pi/T_1=5\pi/3$,
$\epsilon=\sqrt{3}/1.5$, and $h_x=6\omega$.
The sparsely sampled even--odd sequence shows additional frequency
components below the subharmonic frequency $\Omega=0.5$.}
\label{sm:unitgeneric}
\end{figure}

As discussed in the main text, the approximate conservation of
$C_1^x$ and $C_2^x$ restricts the dynamics to configurations with
the same bond correlators as the initial state. The response of
$m^x$ depends on the structure of this sector. In some sectors,
the two bond constraints fix $|m^x|$, and the configurations with
opposite magnetizations can be separated by either a few spin
flips or an extensive number of them. In other sectors, the same
bond constraints allow configurations with different values of
$|m^x|$. Figure~\ref{sm:unitgeneric} illustrates these cases.\\

\paragraph{Bond constraints fix $|m^x|$.}
For a single reversed spin in a fully polarized background, the
bond sector fixes $|m^x|=1-2/L$. Reaching a configuration with
opposite magnetization within this sector requires at least
$L-2$ spin flips. As shown in Fig.~\ref{sm:unitgeneric}(a), the
period-doubled oscillations become more stable over the simulated
time window as $L$ increases.

For a single reversed spin in a N\'eel background, the bond sector
fixes $|m^x|=2/L$. Two spin flips suffice to reach a configuration
with the opposite magnetization in the same sector. The period-doubled
oscillations show a visible modulation [Fig.~\ref{sm:unitgeneric}(b)].\\

\paragraph{Bond constraints do not fix $|m^x|$.}
For two adjacent reversed spins in a fully polarized background,
the bond sector contains configurations with different values
of $|m^x|$. The long-time dynamics of $m^x$ shows modulation of
the period-doubled oscillations, with additional features in
the Fourier spectrum alongside the subharmonic peak at
$\Omega=0.5$ [Fig.~\ref{sm:unitgeneric}(c)].

% \clearpage
% \section*{End Matter}
% \appendix

% \begin{thebibliography}{2}
% \bibitem{Casanova2011}
% J.~Casanova, C.~Sab\'in, J.~Le\'on, I.~L.~Egusquiza, R.~Gerritsma,
% C.~F.~Roos, J.~J.~Garc\'ia-Ripoll, and E.~Solano,
% \href{https://doi.org/10.1103/PhysRevX.1.021018}{Phys. Rev. X \textbf{1}, 021018 (2011)}.
% \bibitem{Zhang2015_Embedding}
% X.~Zhang, Y.~Shen, J.~Zhang, J.~Casanova, L.~Lamata, E.~Solano,
% M.-H.~Yung, J.-N.~Zhang, and K.~Kim,
% \href{https://doi.org/10.1038/ncomms8917}{Nat. Commun. \textbf{6}, 7917 (2015)}.
% \end{thebibliography}
% \end{document}

\end{document}